\documentclass[11pt,twoside]{article}
\usepackage{aussois2003,epsf,psfig,lscape}

\def\edcomment#1{\iffalse\marginpar{\raggedright\sl#1\/}\else\relax\fi}
\reversemarginpar

\def\be{\begin{equation}}
\def\ee{\end{equation}}
\def\bea{\begin{eqnarray}}
\def\eea{\end{eqnarray}}

\def\Real{{\rm I\mathchoice{\kern-0.70mm}{\kern-0.70mm}{\kern-0.65mm}%
  {\kern-0.50mm}R}}

\font \bolditalics = cmmib10
\def\bx#1{\leavevmode\thinspace\hbox{\vrule\vtop{\vbox{\hrule\kern1pt
        \hbox{\vphantom{\tt/}\thinspace{\bf#1}\thinspace}}
      \kern1pt\hrule}\vrule}\thinspace}

\def \vc #1{{\textfont1=\bolditalics \hbox{$\bf#1$}}}

\def\xig{{\vc \xi}}
\def\gammag{{\vc \gamma}}
\def\thetag{{\vc \theta}}
\def\deltag{{\vc \delta}}
\def\varthetag{{\vc \vartheta}}

\def\Sg{{\bf S}}
\def\kg{{\bf k}}
\def\ggr{{\bf g}}
\def\dg{{\bf d}}
\def\eg{{\bf e}}
\def\xg{{\bf x}}
\def\rg{{\bf r}}
\def\sg{{\bf s}}

\begin{document}
%
\title{Gravitational Lensing by Large Scale Structures:\\
A Review}
%
\author{Ludovic Van Waerbeke}
\affil{Institut d'Astrophysique de Paris, 98bis Bd Arago, 75014, Paris}

\author{Yannick Mellier}
\affil{Institut d'Astrophysique de Paris, 98bis Bd Arago, 75014, Paris\\
LERMA, Observatoire de Paris, 61 av. de l'Observatoire, 75014, Paris}
\label{page:first}
\begin{abstract}
We review all the cosmic shear results obtained so far,
with a critical
discussion of the present strengths and 
weaknesses. We discuss the future prospects and the role
cosmic shear could play in a precision cosmology era.
\end{abstract}
\section{Introduction}

The observation of gravitational lensing by large scale structures is a 
direct probe of
the matter distribution in the Universe. This method gives the most
unbiased
picture of the matter distribution at low redshift compared to
other techniques like  cosmic velocity fields, galaxy
distribution or Lyman-$\alpha$ forest studies. Indeed, these techniques
rely on assumptions either like the  
dynamical stage of the structure involved, or the 
properties of visible material versus dark 
matter biasing, or suffer of a poor sampling, 
or a combinaison of those. On the other hand, lensing
by large scale structures suffers from
practical difficulties, like its sensitivity to 
non-linear power spectrum predictions, or to the  Point Spread Function
corrections, which we will discuss later. In this review, we intend to give a 
present day picture of the cosmic shear research and to discuss the
technical issues  that could be a limitation.  These technical limitations
will certainly be overcome sooner or later, this is why a discussion of the 
role of cosmic shear for precision cosmology is also of interest.
Although this paper is supposed to review the topic, there are already more 
than hundreds
of publications on the cosmic shear subject alone. It is therefore  
difficult to address all aspects in details, and to  mention
everything (theory, simulations and observations). Instead,
we choose to focus on observations, data analysis and related  
cosmological interpretations. By {\it cosmic shear}, we mean distorsion
of the distant galaxies only. The magnification aspects of gravitational lensing
by large scale structures, which is only at its beginning in terms of intensive
observations, will not be reviewed.
We apologize whose those of which work will not be discussed.

\section{Linking galaxy shapes to theory}

\subsection{Lensing by large scale structures}

\subsubsection{Light propagation in the inhomogeneous universe}

\begin{figure}
\centerline{\vbox{
\plotone{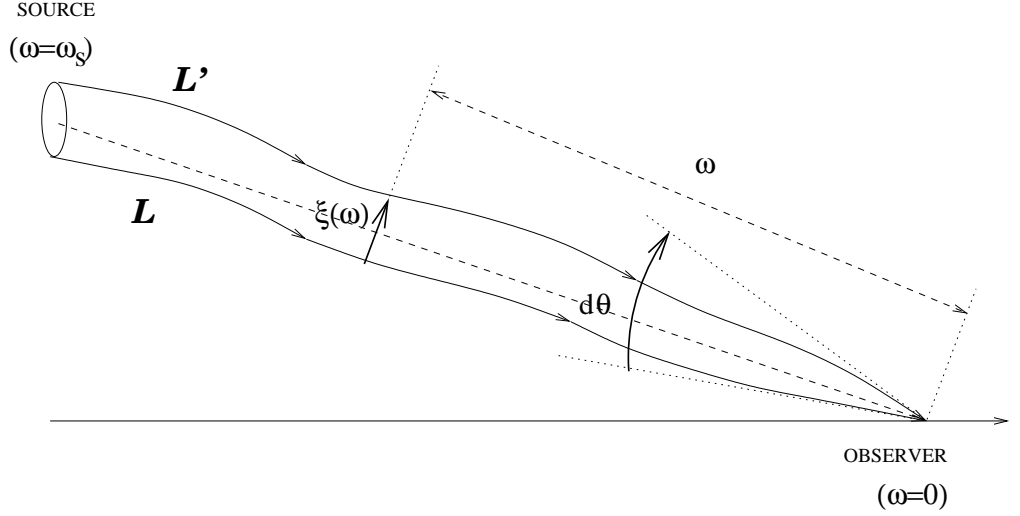} }}
\caption[]{A light bundle and two of its rays $\cal L$ and $\cal L'$. $\xig(w)$ is
the physical diameter distance, which separates the two rays on the sky, viewed from the observer
($w=0$).
\label{inhomo.ps}}
\end{figure}

We first have to define the homogeneous background universe notations (identical
to Schneider et al. 1998). The metric of the
homogeneous Universe is written in the form
\begin{equation}
{\rm d} s^2=c^2\,{\rm d} t^2 -a^2(t)\left[{\rm d} w^2+f_K^2(w) {\rm d}\omega^2\right]\;,
\end{equation}
where $a(t)=(1+z)^{-1}$ is the cosmic scale factor
normalized to unity today, $w(z)$ is the radial coordinate, and $f_K(w)$
is the comoving angular diameter distance out to a distance
$w(z)$. The radial distance $w(z)$ is given by the redshift integral:

\begin{equation}
w(z)=\int_0^z {\rm d} z' {c\over H}={c\over H_0}\int_0^z
{{\rm d} z'\over \sqrt{(1+z')^3 \Omega_0+(1+z')^2 (1-\Omega_0
-\Omega_\Lambda) +\Omega_\Lambda }} \;, 
\end{equation}
where $H_0$ is today's Hubble constant, and the angular diameter distance
$f_K(w)$ reads

\begin{equation}
f_K(w)=\cases{ K^{-1/2} \sin (\sqrt{K} w) & for $K>0$  , \cr
		w				& for $K=0$  , \cr
	(-K)^{-1/2} \sinh (\sqrt{-K} w) & for $K<0$ , \cr}
\end{equation}

where $K$ is the curvature 

\begin{equation}
K=\left({H_0\over c}^2\right) \left(\Omega_0+\Omega_\Lambda -1\right),
\end{equation}
with $\Omega_0$ and $\Omega_\Lambda$ the mean density parameter and the vacuum energy today.

Consider two light rays $\cal L$ and $\cal L'$ coming from a distant source
and converging to an observer, and define ${\rm d}\thetag$ as the observed
angular vector between the two rays (Figure\ref{inhomo.ps}).
We use the Cartesian complex coordinates, so ${\rm d}\thetag=({\rm d}\theta_1,
i{\rm d}\theta_2)$. In the absence of any inhomogeneities along the line of 
sight, the
physical distance between the two rays at an angular distance $f_K(w_S)$ from the observer
to the source is defined as $\xig=f_K(w_S)\;
{\rm d}\thetag$. Due to the inhomogeneities, like clusters of
galaxies, voids and filaments, the physical distance $\xig$ deviates from 
  this simple
relation, and can be linearized as:

\begin{eqnarray}
\xig=f_K(w_S) {\cal A}\; {\rm d}\thetag&=&f_K(w_S)\pmatrix{\kappa+\gamma & -i\omega \cr -i\omega &\kappa-\gamma}\;{\rm d}\thetag\cr
&=&f_K(w_S)(\kappa-i\omega)\;
{\rm d}\thetag+f_K(w_S)\gamma\;{\rm d}\thetag^\star.
\label{xi_def}
\end{eqnarray}
\begin{figure}
\centerline{
\psfig{figure=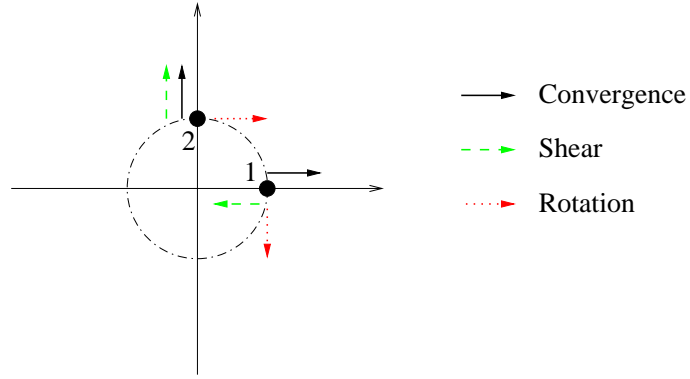,height=5cm}
}
\caption[]{Effect of $\kappa$, $\gamma$ or $\omega$ on the displacement of two test
particles $1$ and $2$ located on a test ring (dot-dashed circle) with coordinates
$({\rm d}\theta,0)$ and $(0,i{\rm d}\theta)$.
\label{matrix_def.ps}}
\end{figure}
The matrix ${\bf {\cal A}}$ is by definition the
amplification matrix. The geometrical origin of this expression is easily
understood when drawing how ${\rm d}\theta_1$ and ${\rm d}\theta_2$ change with a small
(but non-vanishing) $\kappa$, $\gamma$ or $\omega$ (see Figure \ref{matrix_def.ps}).
They are just numbers which describe the
infinitesimal relative displacement of two rays $\cal L$ and $\cal L'$.

\begin{equation}
{\rm d}\xig_\kappa\propto \pmatrix{\kappa\;{\rm d}\theta_1\cr i\kappa\;{\rm d}\theta_2} ; \;
{\rm d}\xig_\gamma\propto \pmatrix{\gamma\;{\rm d}\theta_1\cr -i\gamma\;{\rm d}\theta_2} ; \;
{\rm d}\xig_\omega\propto \pmatrix{\omega\;{\rm d}\theta_2\cr -i\omega\;{\rm d}\theta_1}
\end{equation}

\begin{figure}
\centerline{\vbox{
\psfig{figure=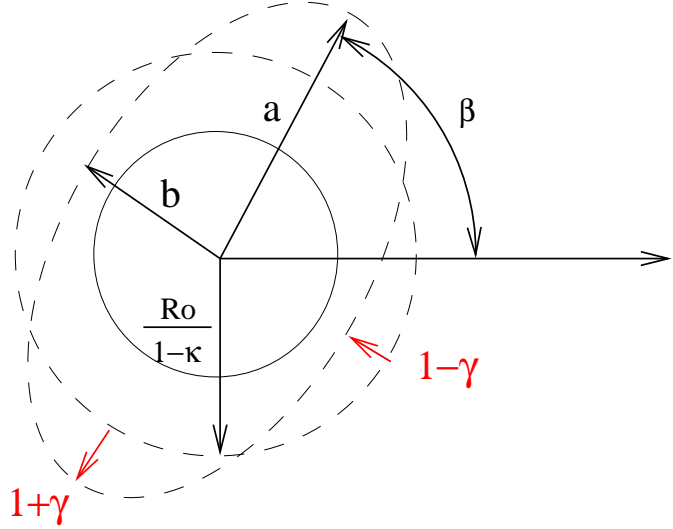,height=7cm}
}}
\caption[]{Illustration of the first order effect of cosmic shear on a circular
background galaxy of radius
$R_0$. The convergence is an isotropic distortion of the image of the galaxy, while the shear is
an anisotropic distortion.
\label{galsection.ps}}
\end{figure}

As we shall see now, they are
the quantities which contain the cosmological information.
By definition, $\kappa$ is called the
convergence field, $\gamma$ the shear field, and $\omega$ the rotation field. When the shear is
not expressed in the eigenspace (which is the case in Figure \ref{matrix_def.ps} for
instance), $\gammag$ is a complex vector in general (see Figure \ref{galsection.ps}).

A light beam is a congruence of null geodesics, which are marked with respect to a fiducial
(reference) geodesic having a tangent vector $k_\mu$. The rays
$\cal L$ and $\cal L'$ are two geodesics of the congruence, whose the separation
$\xig=\xi_1+i\xi_2$ is defined as a space-like vector perpendicular to the wave-vector
$k_\mu$. As above, for an infinitesimal displacement
along the congruence it is always possible to decompose the geometrical deformation
of the ray bundle into a uniform expansion $\Theta$, a shear $\sigma$
and a rotation $W$. This defines the well known optic scalars (Sachs 1961):

\begin{equation}
\Theta={1\over 2} k^\mu_{;\mu} \;\ \ \ \sigma=\sqrt{{1\over 2}[k_{(\mu ;\nu)}k^{\mu ;\nu}-{1\over 2}(k^\mu_{;\mu})^2} \ ; \ \ \ W=\sqrt{{1\over 2}k_{[\mu ;\nu]}k^{\mu ;\nu}},
\end{equation}
where $k_{\mu ;\nu}$ is the covariant derivative of the wave-vector and $(\mu;\nu)$ and $[\mu;\nu]$
denote the symmetric and antisymmetric permutation of indices respectively.
The evolution of the optic scalars along
the congruence is completely determined by the optical scalar equations 
which depend on the gravitational field (Sachs 1961):

\begin{eqnarray}
&&{d(\Theta+iW)\over d\lambda}+(\Theta+iW)^2+|\sigma|^2={\cal R}={1\over 2} R_{\mu\nu}k^\mu k^\nu\nonumber\\
&&{d\sigma\over d\lambda}+\sigma\Theta={\cal F}=C_{\mu\alpha\nu\beta}k^\mu k^\nu \bar t^\alpha \bar t^\beta.
\label{scalar_opt_eqs}
\end{eqnarray}
Here, the geodesic is parametrized with ${\rm d}\lambda={\rm d}w/(1+z)^2$.
$R_{\mu\nu}$ and $C_{\mu\alpha\nu\beta}$ are the Ricci and the Weyl
tensors respectively. $t^\alpha$ is the complex null tetrad (or Sachs tetrad) 
  such that $t^\alpha k_\alpha=0$
and $\bar t_\alpha t^\alpha=1$.
Note that the first equation in (\ref{scalar_opt_eqs}) is nothing else but the Raychaudhuri equation
for null geodesics.

For an infinitesimal displacement along the congruence, the separation $\xig$ 
  transforms according to Eq.(\ref{xi_def}):

\begin{equation}
{d\xig\over d\lambda}=(\Theta-iW)\xig+\sigma\xig^\star.
\label{xi_evol}
\end{equation}
Differentiating Eq.(\ref{xi_evol}) and substituting Eq.(\ref{scalar_opt_eqs})
leads to the evolution equation of $\xig$ along the congruence
as a function of the gravitational fields ${\cal R}$ and ${\cal F}$:

\begin{equation}
{d^2\xig\over d\lambda^2}=\pmatrix{{\cal R}-Re({\cal F}) & iIm({\cal F}) \cr iIm({\cal F}) &
{\cal R}+Re({\cal F})}\xig.
\label{matrix_lens_eq}
\end{equation}
The final step is to calculate $\cal R$ and $\cal F$ from the Ricci and the Weyl tensors for a Newtonian
gravitational potential $\Phi$. Straightforward but lengthy calculations give:

\begin{equation}
{\cal R}=-{1\over a^2(w)}\Delta \Phi \ ; \ \ \ {\cal F}=-{1\over a^2(w)}(\partial_1^2
\Phi-\partial_2^2\Phi+2i\partial_1\partial_2\Phi),
\end{equation}
where $a(w)$ is the scale factor of the unperturbed background metric, and $w$
the radial distance. Using a perturbative expansion for the amplification
matrix ${\cal A}_{ij}={\cal A}^{(0)}_{ij}+{\cal A}^{(1)}_{ij}+...$ and for
the gravitational potential $\Phi=\Phi^{(1)}
+\Phi^{(2)}+...$, Eq.(\ref{matrix_lens_eq}) can be solved iteratively. The homogeneous universe case
corresponds to ${\cal A}_{ij}={\cal A}^{(0)}_{ij}=\delta_{ij}$ and $\Phi=0$.
It is then easy to obtain the general
first order solution for the amplification matrix in the direction $\thetag$:

\begin{equation}
{\cal A}_{ij}(\thetag)=\delta_{ij}+{\cal A}^{(1)}_{ij}(\thetag)=\delta_{ij}-{2\over c^2}\int_0^{w_S} {\rm d}
w\; {f_K(w-w')
f_K(w')\over f_K(w)}\Phi^{(1)}_{,ij}(f_K(w')\thetag,w'),
\label{amplidef}
\end{equation}
where $w_S$ is the position of the source. Eq.(\ref{amplidef})
is the basic lensing equation used to calculate the distortion and the 
   magnification
of distant sources. This result is a first order expression and is only valid 
  in the realm
of the Born approximation where
    the lensing properties are calculated along the unperturbed light path
(of direction $\thetag$).
 Therefore, all contribution coming from the lens-lens coupling 
  are neglected.
For most practical applications this is however an excellent approximation (Bernardeau et al. 1997,
Schneider et al. 1998), as we shall see later.

Back to the lensing effects (Eq.\ref{xi_def}), the geometrical deformation 
   of a light bundle can be expressed as an
integrated effect along the line-of-sight:

\begin{equation}
\kappa=1+{1\over 2}Tr({\cal A}^{(1)}_{ij}) \ ; \ \ \ \gamma={1\over 2}({\cal A}^{(1)}_{11}-{\cal A}^{(1)}_{22}+2i{\cal A}^{(1)}_{12})
 \ ; \ \ \ \omega=0.
 \label{kappagdef}
\end{equation}
These expressions show that a scalar perturbation will never induce a rotation of the light bundle
at the first order ($\omega=0$). Figure \ref{galsection.ps} shows the effect of cosmic shear on a distant
circular galaxy, at the first order ($\kappa\;\ll\;1$ and $\gamma\;\ll\;1$). It shows that the shear
can be obtained from the measurement of the shape of galaxies. The practical methods to do this measurement
will be discussed in Section 2.2.

\subsubsection{Mean fields}

The second order derivatives of the gravitational potential field can be written as
function of the mass density contrast $\delta$, using the Poisson
equation:

\begin{equation}
\nabla^2\Phi={3 H_0^2 \Omega_{\rm 0}\over 2 a}\delta.
\end{equation}
From Eq(\ref{amplidef}), we get the convergence $\kappa(\thetag)$ in the direction $\thetag$,
as function of $\delta$, integrated along the line of sight:

\begin{equation}
\kappa(\thetag,w)={3\over 2}{H_0\over c}^2\Omega_{\rm 0}
\int_0^{w_S}
{\rm d} w'\;{f_K(w-w')\,f_K(w')\over f_K(w)}
{\delta\left(f_K(w')\vc\theta,w'\right)\over a(w')},
\label{kappaeq}
\end{equation}
with  similar (but not identical) expressions for $\gammag(\thetag)$. 
 The sources have been assumed to be at a single 'redshift' $w_S$,
 but similar expressions can be easily 
  generalized for a more realistic redshift distribution. In that case, the
lensing fields are integrated along the redshift with the proper source
distribution $p_w(w){\rm d}w$ from $0$ to the horizon $w_{\rm H}$:

\begin{equation}
\kappa(\thetag)=
{3\over 2}\left(H_0\over c\right)^2\Omega_{\rm 0}
\int_0^{w_{\rm H}} {\rm d} w\;
g(w)\,f_K(w){\delta\left(f_K(w)\vc\theta,w\right)\over a(w)},
\label{kappadef}
\end{equation}
with

\begin{equation}
g(w)=\int_w^{w_{\rm H}}{\rm d} w'\;p_w(w')\,{f_K(w'-w)\over f_K(w')}.
\label{gfunc}
\end{equation}

\subsubsection{Limber equation and small angle approximation}

We are primarily interested in the statistical properties of the lensing 
fields, which are given
by the moments of the field. The variance is the first non trivial moment;  
 its evolution with angular scale depends on cosmological parameters 
  and on the geometrical
properties of the Universe due to the light rays propagation.
The mass density power spectrum $P_{3D}(k)$ is defined as

\begin{equation}
\langle\tilde\delta(\kg)\tilde\delta^*(\kg')\rangle=
(2\pi)^3 \delta_{\rm D}(\kg-\kg')\,P_{3D}(k,w).
\label{p3d}
\end{equation}
Likewise, one can define the convergence power spectrum $P_{\kappa}(s)$:

\begin{equation}
\langle\tilde\kappa(\vc s) \tilde\kappa^*(\vc s')\rangle
=(2\pi)^2 \delta_{\rm D}(\sg-\sg')\,P_\kappa(s).
\label{kappapower}
\end{equation}
The time dependence in Eq(\ref{p3d}) stands for the growth of structures. 
 For an EdS Universe,
it can be factorized, but in the general case it is more complicated, 
 in particular in the
non-linear regime where time dependence and scales are coupled.
The jump from the 3-D wave vector $\kg$ to the 2-D angular wave vector 
$\sg$ is ensured from the
line of sight integration using the Limber approximation (Limber, 1954). 
To simplify
Eq(\ref{kappadef}), it
can be written as $\kappa(\thetag)=\int{\rm d}w\; q(w) \; \delta(f_K(w)\vc\theta,w)$.
In real space,
 the convergence correlation function $\xi_\kappa(\Delta\thetag)=
\langle\kappa(\thetag)\kappa(\thetag+\Delta\thetag)\rangle$
 can be eventually computed (Kaiser 1998):

\begin{eqnarray}
\langle\kappa(\thetag)\kappa(\thetag+\Delta\thetag)\rangle&=&
\int{\rm d}w\; q(w) \int{\rm d}w'\; q(w') \; \langle\delta(f_K(w)\vc\theta,w)
\delta(f_K(w')(\vc\theta+\Delta\thetag),w')\rangle \cr
&\simeq &
\int{\rm d}w\; q^2(w) \int{\rm d}w' \; \langle\delta(f_K(w)\vc\theta,w)
\delta(f_K(w')(\vc\theta+\Delta\thetag),w')\rangle,
\label{kappacorr}
\end{eqnarray}
 assuming that the selection function 
$q(w)$ does not vary
across the largest fluctuations of the density and that the fluctuations 
  are much smaller than the distance of the sources.
In order to express all cosmic shear 2-points statistics, we are in fact 
   interested in the convergence power spectrum $P_\kappa(s)$:

\begin{equation}
P_\kappa(s)=\int{\rm d}\thetag\;\xi_\kappa(\thetag)\;{\rm e}^{-i\sg\cdot\thetag}.
\end{equation}
The density contrast $\delta(f_K(w)\vc\theta,w)=\delta(\rg)$ can be expressed in Fourier
space:

\begin{eqnarray}
\delta(\rg)&=&\int{{\rm d}\kg\over (2\pi)^3}\;{\rm e}^{-i\kg\cdot\rg}\;\tilde\delta(\kg,w)\cr
&=&\int{{\rm d}\kg\over (2\pi)^3}\;{\rm e}^{-i\kg_\perp\cdot\thetag\;f_K(w)}\;{\rm e}^{-i\; k_3 w}\;D_1^{(+)}(w)\;
\tilde\delta(\kg),
\end{eqnarray}
where $D_1^{(+)}(w)$ is the linear structure growth factor 
(see the next section
{\it non-linear power spectrum}), and $\kg=(\kg_\perp,k_3)$, $\kg_\perp$ is
the wave-vector perpendicular to the line of sight. From this equation and 
 Eq(\ref{p3d}),
one can express the density correlation function appearing in Eq(\ref{kappacorr}):

\begin{eqnarray}
\langle\delta(\rg)\delta^\star(\rg')\rangle&=&\int{\rm d}\kg\;
{\rm e}^{-i\kg_\perp\cdot\thetag\;f_K(w)}\;
{\rm e}^{i\kg_\perp\cdot(\thetag+\Delta\thetag)\;f_K(w')} \cr
&\times& {\rm e}^{-i\;k_3 (w-w')}\;D_1^{(+)}(w)D_1^{(+)}(w')\;P_{3D}(k).
\end{eqnarray}
When, as in our case, the small angle approximation is 
  valid ( $|\Delta\thetag| \le 1-2\;{\rm degrees}$),
 the transverse wave-vector $\kg_\perp$
carries most of the power at $|\kg|$; that is 
 $P_{3D}(k)\simeq P_{3D}(k_\perp)$ (Peebles 1980).
The $k_3$ integration then gives a Dirac delta 
   function $\delta_D(w-w')$.
If we perform the variable change $\kg_\perp\;f_K(w)=\sg$ the convergence power spectrum
becomes:

\begin{equation}
P_\kappa(s)=\int{\rm d}w\;{q^2(w)\over f_K^2(w)}\left[D_1^{(+)}(w)\right]^2\;
P_{3D}\left({s\over f_K(w)}\right).
\end{equation}
Back to the notations of Eq(\ref{kappadef}), the convergence power spectrum 
    finally writes

\begin{equation}
P_\kappa(s) =
{9\over 4}\left(H_0\over c\right)^4\Omega_{\rm 0}^2
\int_0^{w_{\rm H}}{\rm d} w\;{g^2(w)\over a^2(w)} 
P_{3D}\left({s\over f_K(w)};w\right).
\label{pofkappa}
\end{equation}
The shear power spectrum $P_\gamma(s)$ is identical to this expression. The 
  reason is that,
in Fourier space, the quantities $\langle\tilde \kappa^2\rangle$ 
 and $\langle|\tilde\gammag|^2\rangle$ are identical.
This can be derived easily from Eq(\ref{amplidef}) and Eq(\ref{kappagdef}), 
  with the derivatives
replaced by powers in $\sg$'s in Fourier space. As we shall see, this allows us to
extract the convergence 2-points statistics directly from the data. Higher order
statistics is a more difficult issue which will be discussed later.

\subsubsection{Non-linear power spectrum}

The normalization of the mass density power spectrum $P_{3D}$ is defined 
   in the
conventional way, by computing the mass density variance within a sphere of
$8\;{\rm h}^{-1}{\rm Mpc}$ radius at redshift zero:

\begin{equation}
\sigma_8^2=\langle \delta_R^2\rangle={1\over 2\pi^3}\int d\kg P_{3D}(k,0) |W(kR)|^2,
\label{sigma8}
\end{equation}
where $W(kR)={3\over (kR)^2}\left({\sin (kR)\over kR}-\cos (kR)\right)$ 
 is the Fourier
transform of the top-hat window function of radius $R$. The transition from 
   the
linear to the non-linear scales is identified by $\sigma_8\sim 1$.  
In the linear regime, where the density contrast of the mass distribution is 
  low
($\delta\; \ll\; 1$), the fluid equations describing the structure growth can
be solved perturbatively, and one obtains for the growing mode:

\begin{equation}
P_{3D}(k,w)=\left[D^{(+)}_1(w)\right]^2\;P_{3D}(k),
\end{equation}
with,

\begin{equation}
D^{(+)}_1(w)={5\over 2}\;\Omega_0\;H(w)\;\int_0^w\;{{\rm d}a\over a^3\;H(a)}.
\end{equation}
\begin{figure}
\centerline{\vbox{
\psfig{figure=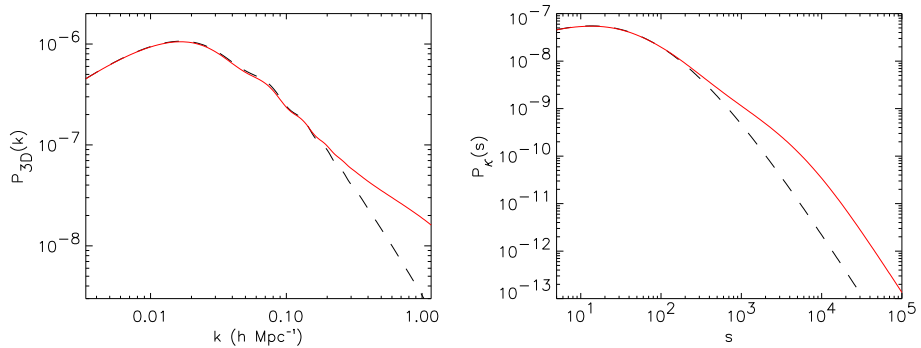,height=5cm}
}}
\caption[]{The left panel is a 3-dimensional mass power spectrum for the
linear (dashed) and non-linear (solid, using Smith et al. 2002) regimes
when baryons are included.
A value of $\Omega_b=0.05$
was used. The right panel shows the induced convergence power spectrum
(Eq.\ref{pofkappa}) for the two dynamical regimes. Other parameters
are $\Omega_{\rm cdm}=0.25, \Omega_\Lambda=0.7, \sigma_8=0.9, h=0.7,
z_{\rm source}=0.8$.
\label{powernl.ps}}
\end{figure}
In the non-linear regime, the structure growth cannot be solved analytically 
   and its description must  rely on  non-linear models 
  (Peacock \& Dodds, 1996, Smith et al. 2002), following
an original idea of Hamilton et al. (1991). Non-linear predictions of the matter
power spectrum are performed from the knowledge of the
spatial 2-points correlation function of the galaxies
$\xi_2(\rg)={V\over (2\pi)^3}\int\;{\rm d}\kg\;P(k)\;e^{-i\kg\cdot\rg}$. An
accurate measurement of $\xi_2(\rg)$ is given for instance by the 
   2dF (Percival et al. 2001) or the SDSS surveys (Dodelson, S., et al. 2002):

\begin{equation}
\xi_2(r)=\left({r_0\over r}\right)^\gamma,
\end{equation}
with $r_0=4.3\pm0.3\;h^{-1}{\rm Mpc}$ and $\gamma=1.71\pm0.06$. The
stable clustering hypothesis stipulates that at very small scales (strong
non-linear regime), the internal profile of clusters of galaxies
remain constant with time for any cosmological model, and that the cluster
distribution is driven by the cosmic expansion.
This means that the correlation function is fixed in proper coordinates, but
its amplitude evolves as a volume effect like $(1+z)^{-3}$. At large scale
(linear regime), the correlation function follows the perturbation theory.
Since the correlation function $\xi_2(r)$ behaves like
$r^{-\gamma}$ for any cosmological model, we therefore have the two following
limiting cases (Peacock 1999):

\begin{equation}
\xi_2(r,z)\propto (1+z)^\gamma (1+z)^{-3} \ \ \ {\rm non-linear}
\end{equation}

\begin{equation}
\xi_2(r,z)\propto \left[D^{(+)}_1(w)\right]^2  \ \ \ {\rm linear}
\end{equation}
A mapping from the linear to the non-linear
scale has been conjectured (Hamilton et al. 1991, Peacock \& Dodds 1996,
Smith et al. 2002),
and calibrated using N-body simulation. The transition 
   from
linear to non-linear scales is described by a few slowly varying functions 
 that depend on cosmological parameters. The same argument applies to
the 3-D power spectrum, which is needed for cosmic shear predictions
down to small scales
(Eq.\ref{pofkappa}) (Peacock \& Dodds 1994).
Figure \ref{powernl.ps} is an example of 3-dimensional and convergence power
spectra in the linear and non-linear regimes. A fair amount of baryons
was included (using CAMB,
Lewis et al. 2002), in order to show that the baryon oscillations, which
are clearly visible on the 3D spectrum, are severely diluted in the
projected spectrum.

\subsubsection{2-points statistics}
\begin{figure}
\centerline{\vbox{
\psfig{figure=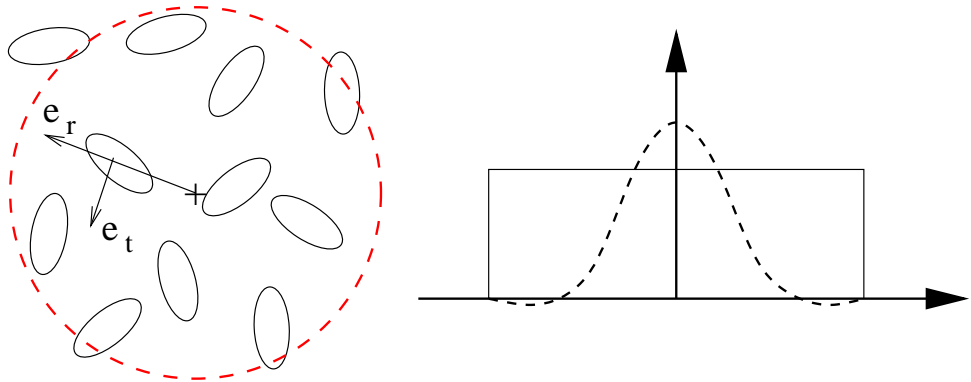,height=5cm}
}}
\caption[]{In order to compute the shear variances, the galaxy
ellipticities are smoothed within a window (dashed red) of fixed radius
$\theta_c$ (left). The shear variance will show up as an excess of galaxy
alignment with respect to random orientation. The right panel shows the
profile of the two filters one usually consider, top-hat (solid line) and
compensated (dashed line). On the left, the axis $(e_t,e_r)$ correspond
to the local frame attached
to each individual galaxy, on which the galaxy ellipticity components can
be projected out to give an estimate of the tangential $\gamma_t$
and radial shear $\gamma_r$.
\label{filter.eps}}
\end{figure}

In practice,  the variance of the convergence (or shear, which is 
  the same) is computed within a
given smoothing window $U(\thetag)$ of radius $\theta_c$, which can be written:

\begin{eqnarray}
\langle \kappa^2\rangle_{\theta_c} &=& \langle\left(\int{\rm d}^2\theta'\;
U(\theta')\kappa(\theta')\right)^2\rangle\cr
&=& \int{\rm d}^2\theta'\;U(\theta')\int{\rm d}^2\vartheta\;U(\vartheta)
\langle\kappa(\thetag')\kappa(\varthetag)\rangle.
\end{eqnarray}
If we express the convergence from its Fourier transform $\kappa (\thetag)=\int {\rm d}^2\sg \;\tilde
\kappa (\sg) \;{\rm e}^{i\thetag\cdot\sg}$ and using Eq(\ref{kappapower}), we 
  obtain:

\begin{eqnarray}
\langle \kappa^2\rangle_{\theta_c}&=&\int{\rm d}^2\theta'\;U(\theta')\int{\rm d}^2\vartheta\;U(\vartheta)
\int{{\rm d}^2 s\over (2\pi)^2}\;{\rm e}^{i\sg
\cdot(\vc\theta'-\varthetag)} P_\kappa(s) \cr
&=&2\pi \int_0^\infty {\rm d} s\;s\,P_\kappa(s)
\left(\int_0^{\theta_c}{\rm d}\vartheta\;\vartheta\,U(\vartheta)\,{\rm J}_0(s\vartheta)\right)^2.
\end{eqnarray}
This expression is general, and can be applied to any smoothing 
  window $U(\thetag)$. 
 Since $P_\gamma(s)=P_\kappa(s)$,
 it also    
expresses the shear variance $\langle\gamma^2\rangle_{\theta_c}$. 
As illustrated in Figure \ref{filter.eps} , we are primarily 
   interested in a top-hat filtering, for which,

\begin{equation}
\langle\gamma^2\rangle={2\over\pi}\int{\rm d}s\;s\;P_\kappa(s)\left[{J_1(s\theta_c)\over s\theta_c}\right]^2,
\label{tophatstat}
\end{equation}
and in the  compensated filtering having  
  $\int_0^{\theta_c}{\rm d}\theta\;\theta\;U(\theta)=0$
(zero mean).
The choice of $U(\theta)$ is arbitrary, provided it has a zero mean. 
  Here we use
the expression (Schneider et al. 1998):

\begin{equation}
U(\theta)={9\over \pi\theta_c^2}\left(1-\left({\theta\over\theta_c}\right)^2\right)
\left({1\over 3}-\left({\theta\over\theta_c}\right)^2\right),
\end{equation}
 so the variance of the convergence with this filter is:

\begin{equation}
\langle M_{\rm ap}^2\rangle={288\over\pi}\int{\rm d}s\;s\;P_\kappa(s)\left[{J_4(s\theta_c)\over s^2\theta_c^2}\right]^2.
\label{mapstat}
\end{equation}
\begin{figure}
\centerline{\vbox{
\psfig{figure=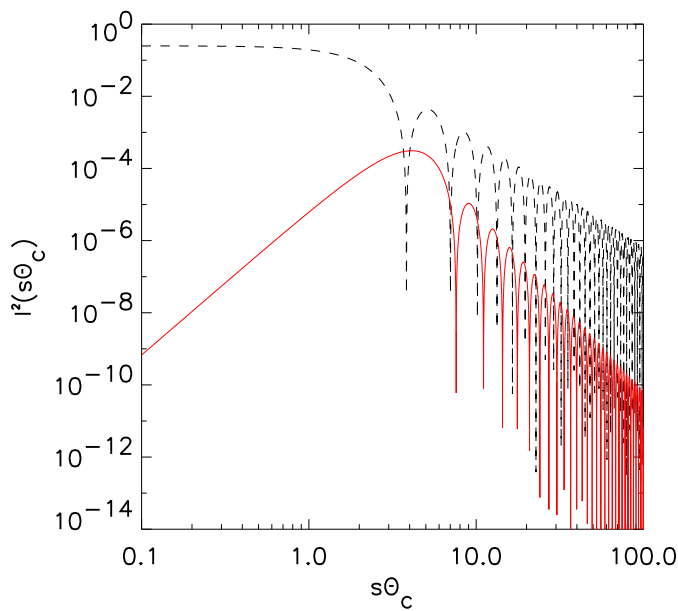,height=8cm}
}}
\caption[]{Top-hat (dashed line) and compensated (solid line) filters in Fourier
space. This plot illustrates the fact that the compensated filter is a pass-band
filter, and therefore is a broad-band estimates of the convergence power spectrum
in real space.
\label{fourierfilter.ps}}
\end{figure}
The nice feature of the  compensated filter is that it is a 
  pass-band filter, which means
that the variance Eq(\ref{mapstat}) is a direct estimate of the convergence 
   power
spectrum in real space. Note that the power is estimated around 
   $s\sim \;5/\theta_c$.
 Furthermore,  it can
be estimated directly from the ellipticity of the galaxies, without a 
   reconstruction
of the convergence field. This  remarkable property 
   has been demonstrated by Kaiser et al. (1994), who
have shown that Eq(\ref{mapstat}) can be obtained from a smoothing of the 
   tangential
component of the shear field $\gamma_t$:

\begin{equation}
M_{\rm ap}=\int_0^{\theta_c} {\rm d}\thetag\;Q(\theta)\;\gamma_t,
\end{equation}
where

\begin{equation}
Q(\theta)={2\over \theta_c^2}\int_0^{\theta_c}{\rm d} \theta'\; \theta'\,U(\theta')
-U(\theta).
\end{equation}
The tangential shear $\gamma_t$ can be obtained from the projection of 
   the galaxy
ellipticity on the local frame
(Figure \ref{filter.eps}).

\begin{figure}[t]
\centerline{\vbox{
\psfig{figure=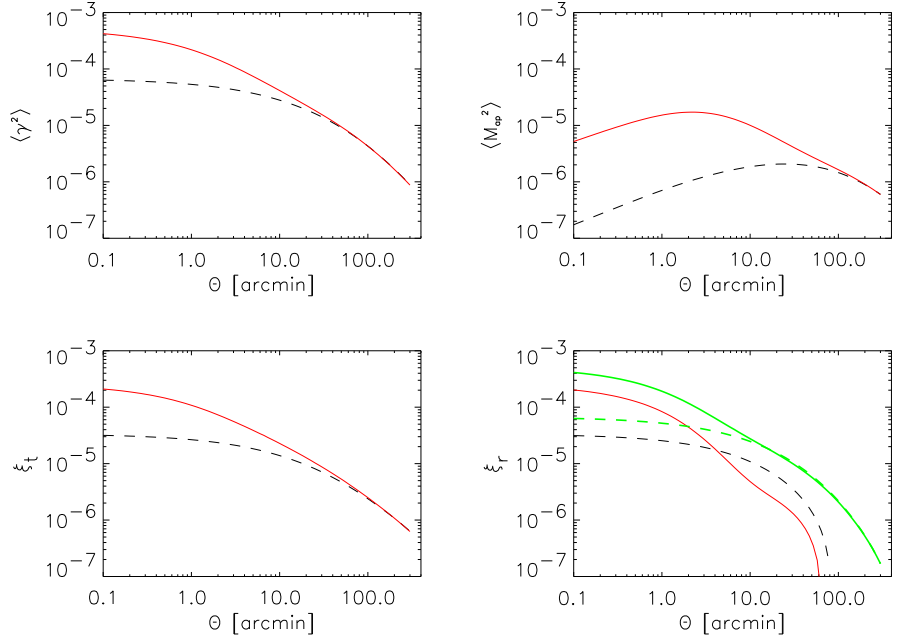,height=9cm}
}}
\caption[]{Lensing statistics predictions for the cosmological model used in
Figure \ref{powernl.ps}. Both linear (dashed) and non-linear (solid lines) regimes
are represented. On the bottom-right plot, the thick dashed and solid lines
are the full shear correlation function. The cosmological model is the same
as Figure \ref{powernl.ps}.
\label{stats.ps}}
\end{figure}
Another 2-points statistics of interest is the shear correlation function
$\langle \gamma\cdot\gamma\rangle_{\theta_c}$. It consists in calculating the 
  sum
of the shear product of all possible galaxy pairs separated by a 
  distance $\theta_c$.
Using the shear field version (i.e. for $\gammag$)
of Eq(\ref{kappaeq}), one can show that (Blandford et al. 1991,
Miralda-Escude 1991, Kaiser 1992):

\begin{equation}
\langle\gammag\cdot\gammag\rangle_{\theta_c}={1\over 2\pi}\int{\rm d}s\;s\;P_\kappa(s)\;
J_0(s\theta_c).
\label{ggstat}
\end{equation}
One can also compute the shear correlation functions 
  of the projected components of the shear,
$\langle\gamma_t\;\gamma_t\rangle$, $\langle\gamma_r\;\gamma_r\rangle$. 
  For
symmetry reasons $\langle\gamma_t\;\gamma_r\rangle=0$. On the other hand, the 
  two
former correlation functions are not equal, because the
gravitational shear is generated by a scalar potential, implying that the 
  projections
on the local frame of the shear components are not equivalent. We can 
  show that:

\begin{eqnarray}
\langle\gamma_t\;\gamma_t\rangle_{\theta_c}&=&{1\over 4\pi}\int{\rm d}s\;s\;P_\kappa(s)\;
\left[J_0(s \theta_c)+J_4(s \theta_c)\right] \cr
\langle\gamma_r\;\gamma_r\rangle_{\theta_c}&=&{1\over 4\pi}\int{\rm d}s\;s\;P_\kappa(s)\;
\left[J_0(s \theta_c)-J_4(s \theta_c) \right]
\label{corrfuncdef}
\label{etetererstat}
\end{eqnarray}
One usually denotes $\xi_+(\theta_c)=\langle\gamma_t\;\gamma_t\rangle+
\langle\gamma_r\;\gamma_r\rangle$,
and 
 $\xi_-(\theta_c)=\langle\gamma_t\;\gamma_t\rangle-\langle\gamma_r\;\gamma_r\rangle$. We
have, of course, $\xi_+(\theta_c)=\langle\gammag\cdot\gammag\rangle_{\theta_c}$.

Figure \ref{stats.ps} shows the linear and non-linear predictions for all the
statistics defined here, for a particular cosmological model.

\subsubsection {Dependence on cosmological parameters}
\begin{figure}[t]
\centerline{\vbox{
\psfig{figure=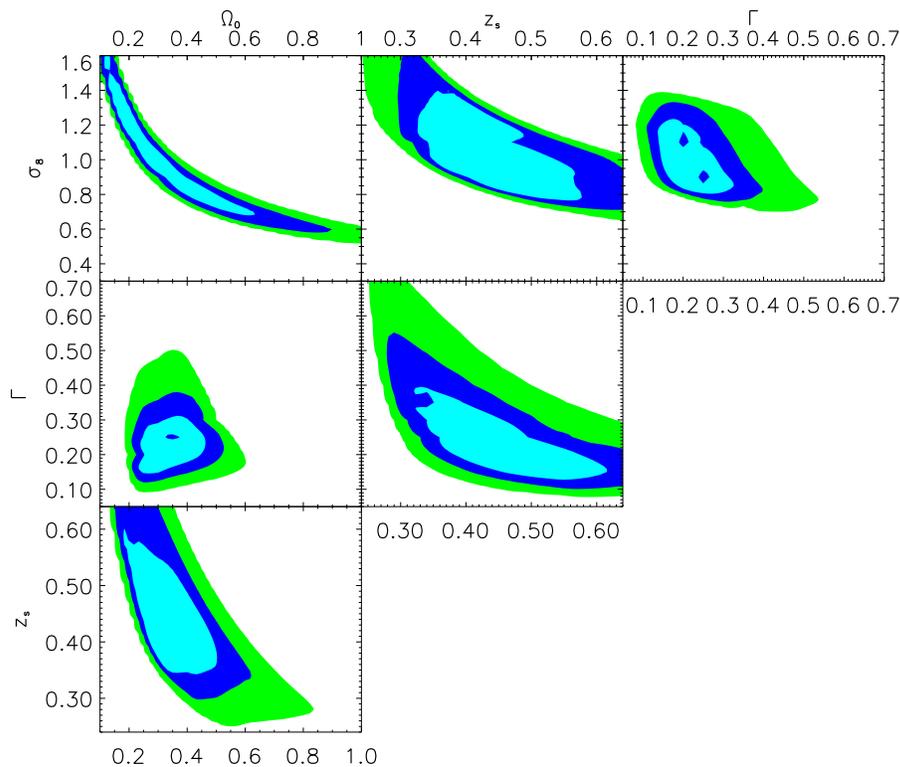,height=10cm}
}}
\caption[]{
1-$\sigma$,
   2-$\sigma$ and 3-$\sigma$ confidence contours for the maximum
   likelihood analysis on the four parameters $\Omega_{\rm m}$,
   $\sigma_8$, $\Gamma$ and the source redshift parameter $z_{\rm s}$
   (see text). The six possible pairs of parameters are displayed. On
   each figure, the two hidden parameters are marginalized such that
   $\Omega_{\rm m}\in [0.2,0.4]$, $\sigma_8\in [0.8,1.1]$, $\Gamma\in
   [0.1,0.3]$ and $z_{\rm s}\in [0.4,0.5]$, and the cosmological
   constant is fixed to $\Omega_\Lambda=1-\Omega_{\rm m}$. The reference
   model is $\Omega_{\rm m}=0.3$, $\sigma_8=1$, $\Gamma=0.21$ and
   $z_{\rm s}=0.44$. The survey area is $A=16\,{\rm deg}^2$, the
   galaxy ellipticity r.m.s. is $0.3$, and the correlation functions
   are measured in the range $0'6< \theta < 30'$.
\label{weakprior.ps}}
\end{figure}
\begin{figure}[t]
\centerline{\vbox{
\psfig{figure=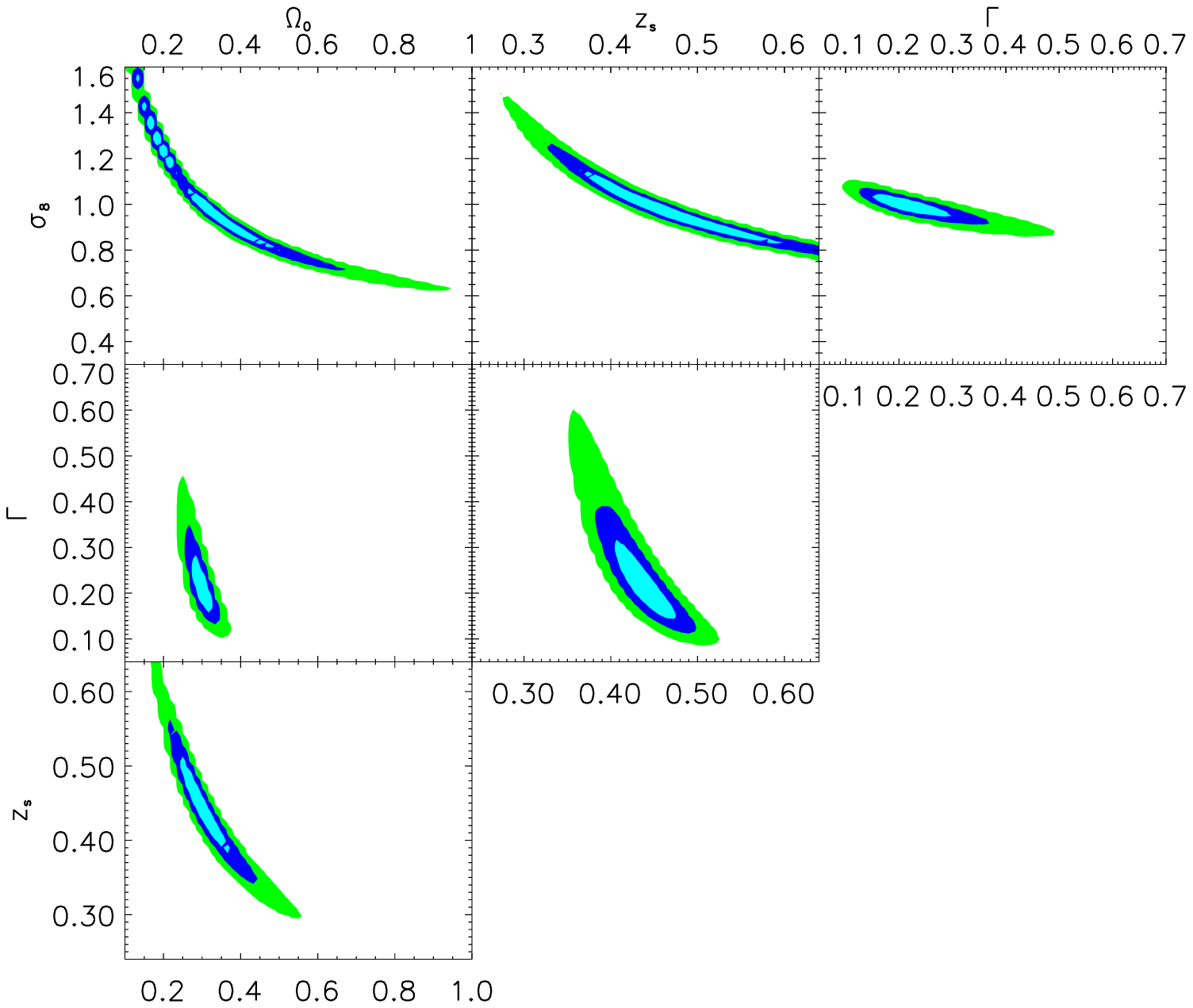,height=10cm}
}}
\caption[]{Same as
   figure \ref{weakprior.ps} with strong priors: in each figure, the
   two hidden parameters as assumed to be known perfectly. These plots
   show the degeneracy directions among all the possible pairs of
   parameters obtained from $\Omega_{\rm m}$, $\sigma_8$, $\Gamma$ and
   $z_{\rm s}$.
\label{strongprior.ps}}
\end{figure}

It is obvious from Eq(\ref{kappaeq}), Eq(\ref{pofkappa}) and Eq(\ref{sigma8}) 
  that the
cosmic shear signal depends primarily on the source redshift $w_S$,
then on the 
   mean density parameter $\Omega_0$, and on 
   the slope and the normalization ($\sigma_8$)
of the mass power spectrum. To explore the parameter dependence of the cosmic 
  shear
signal, we assume the Cold Dark Matter model, with a power spectrum 
  parameterized
with the slope parameter $\Gamma$. We allow the four parameters
($\Omega_0, z_s, \Gamma, \sigma_8$) to vary, and we compute the likelihood
${\cal L}(\Omega_0, z_s, \Gamma, \sigma_8\;|\;{\rm\bf d})$
of the parameters knowing the data $\bf d$. The data vector is for instance the
aperture mass or any other statistic:

\begin{equation}
{\cal L}={1\over (2\pi)^{n/2} \left|\Sg\right|^{1/2}}~\exp{\left[-{1\over 2}\left({\rm \dg-\sg}
\right)^T \Sg^{-1}\left({\rm \dg-\sg}\right)\right]},
\label{likelihood}
\end{equation}
where $\sg$ is the fiducial model vector and
$\Sg:=\langle\left({\rm \dg-\sg}\right)^T\left({\rm \dg-\sg}\right)\rangle$
is the covariance matrix.
Figure \ref{weakprior.ps} and \ref{strongprior.ps} show the parameter 
   dependence
one expects for a survey covering $16$ square degrees up to the limiting
  magnitude $I_{AB}=24$, for two different choices of priors.
The signal also depends on other cosmological parameters
($\Omega_b$, $\Omega_\Lambda$, $\Omega_\nu$,...), albeit to 
  a lower extend. For precision cosmology, all parameters
are relevant, but the first constraints obtained so far from cosmic shear are
on the main four parameters ($\Omega_0, z_s, \Gamma, \sigma_8$).

\subsection{Galaxy ellipticities and estimators}

\subsubsection{Ellipticity of the galaxies}

As mentioned in the previous Section, the cosmic shear signal is measured 
  from the shape
of the distant lensed galaxies. It is quantified from  the ellipticity 
   $\eg$.
The raw ellipticity
$\eg$ of a galaxy is measured from the second moments $I_{ij}$ of the surface
brightness $f(\thetag)$:

\begin{equation}
\eg=\left({I_{11}-I_{22}\over Tr(I)} ; {2I_{12}\over Tr(I)}\right), \ \ \
I_{ij}=\int {\rm d}^2\theta
W(\theta)\theta_i\theta_j f(\thetag).
\end{equation}
The window function $W(\thetag)$ suppresses the noise at large distances
from the object center. The cosmic shear signal can also be measured 
  using gravitational magnification from the
relative size and number count of the lensed galaxies, but this is out of the 
   scope
of this paper. Here, we only focus on the gravitational distortion effect.
If one could measure the shape of the galaxies (with $W(\thetag)=1$)
perfectly without any systematics
coming from the telescope tracking and the optical defects, and if the galaxies
were {\it only} lensed, then the observed ellipticity would be related to the source
ellipticity as

\begin{equation}
\eg^{obs}={\eg^{source}+\ggr\over 1+\eg^{source}\cdot\ggr},
\label{distodef}
\end{equation}
where $\ggr=\gammag/(1-\kappa)$ is the reduced shear, and $\eg^{obs}$ is
the observed ellipticity, $\eg^{source}$ is the source (unobserved)
ellipticity. For nearly all cosmic shear
application, the lens fields are small ($|\ggr |,\kappa\ll 1$) and the linear approximation
is valid $\eg^{obs}\simeq\eg^{source}+\gammag$.

\begin{figure}[t]
\centerline{
\psfig{figure=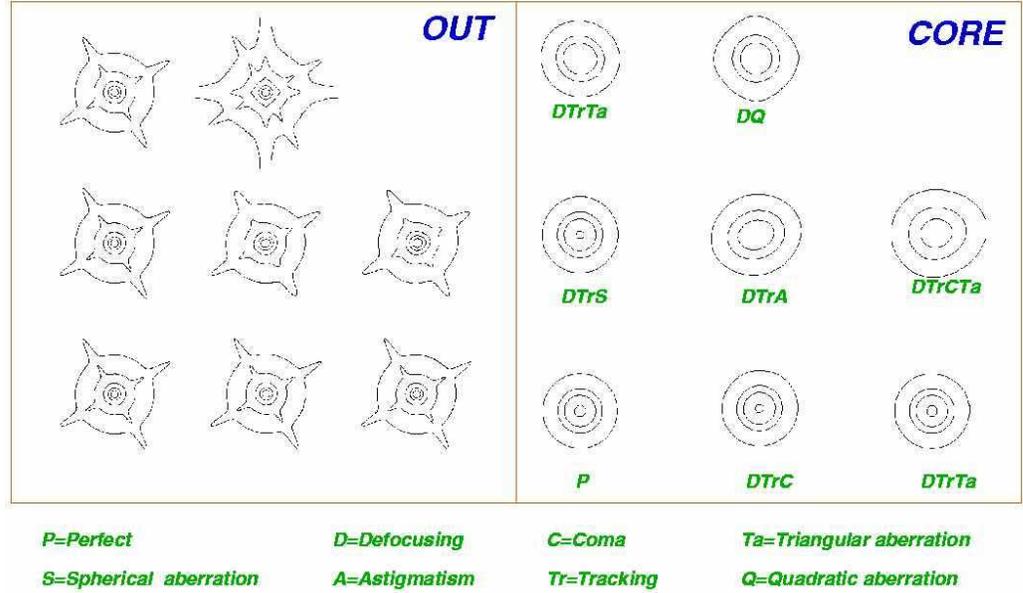,height=8cm}
}
\caption[]{Right plots: simulated cores of anisotropic PSF's. Left plots: simulated
outer part of the PSF (Erben et al. 2001). The different PSFs are computed from
ray-tracing through the telescope optic.
\label{psf.ps}}
\end{figure}
\begin{figure}[t]
\centerline{
\psfig{figure=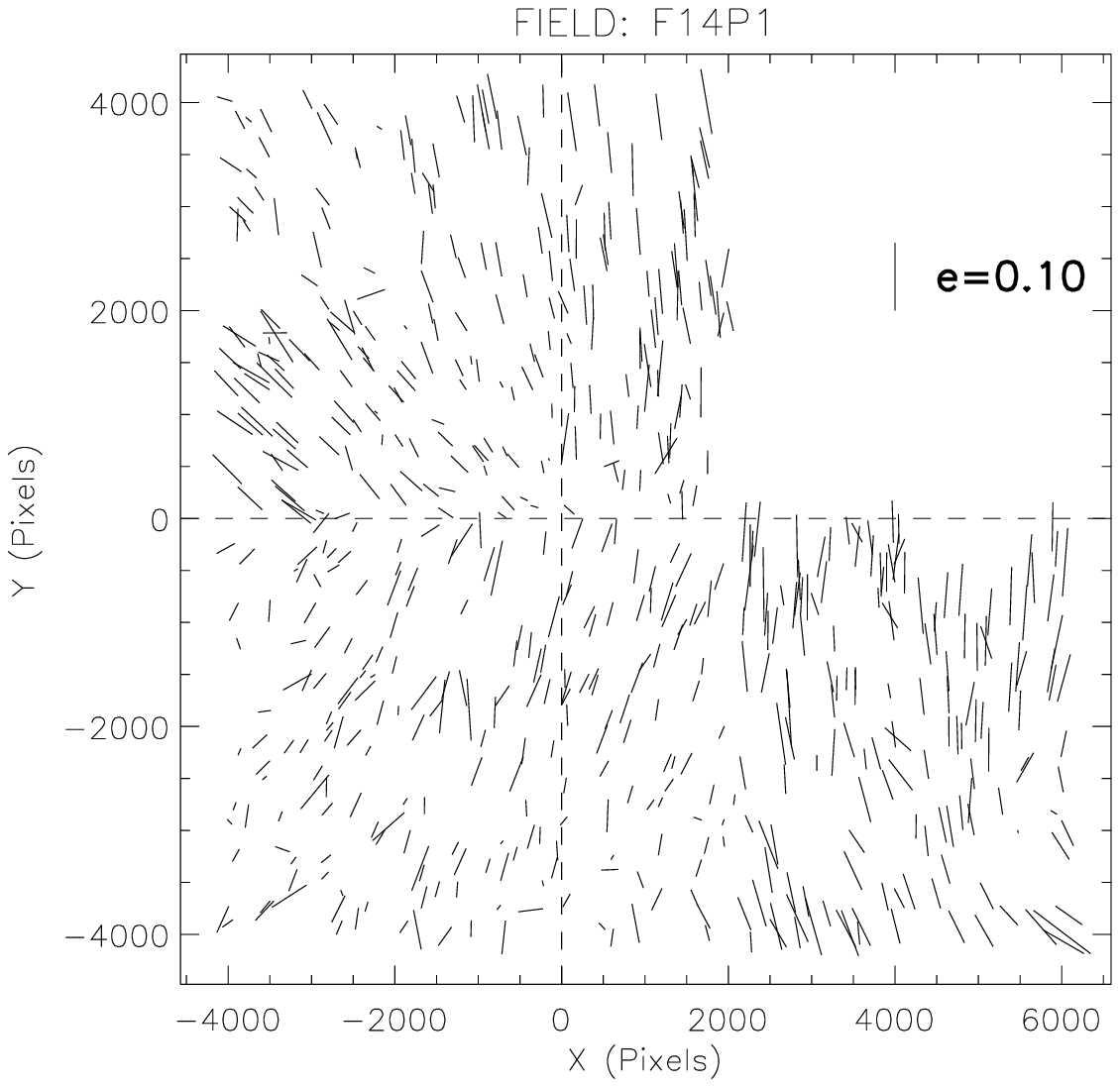,height=7cm}
\psfig{figure=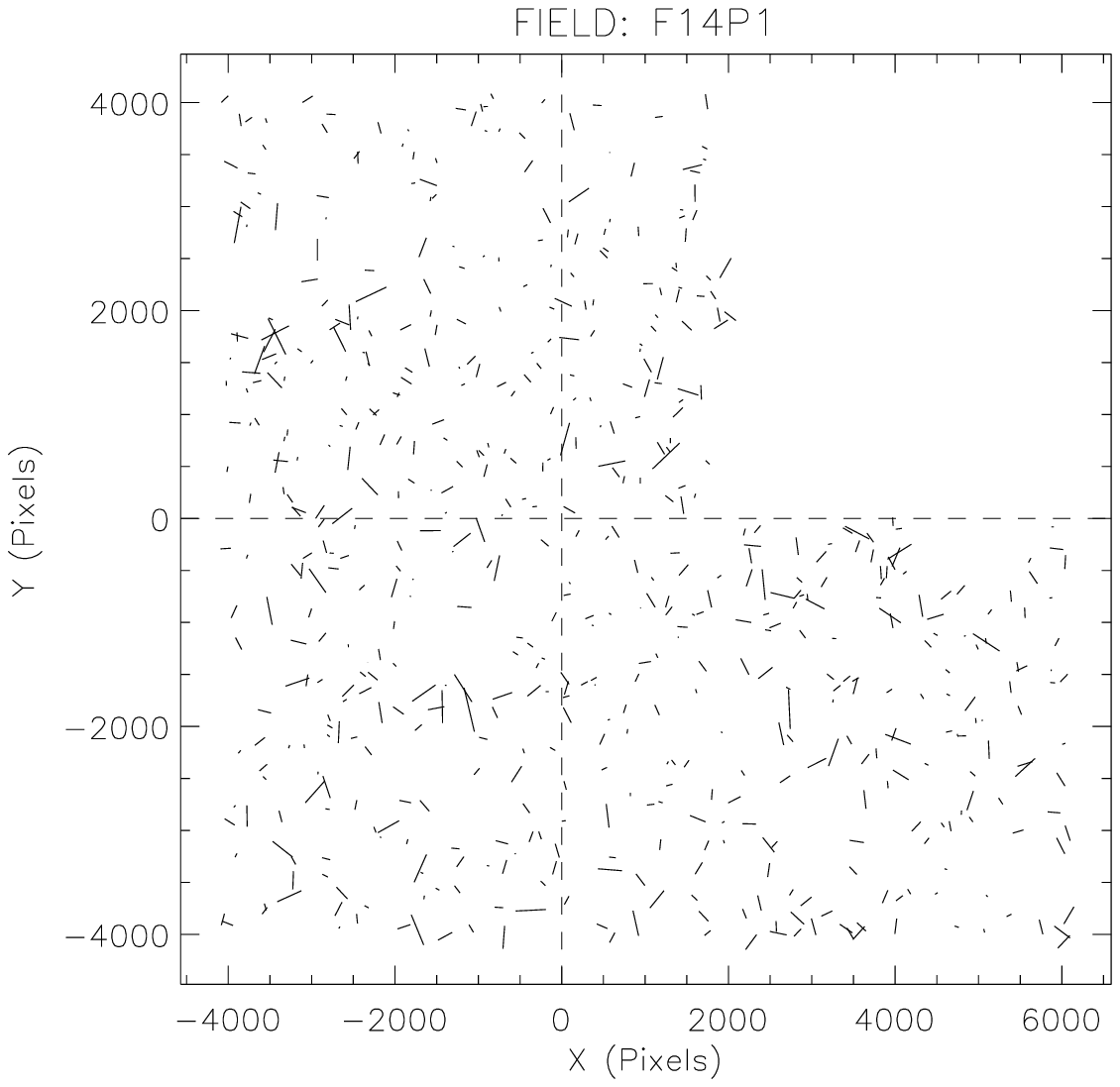,height=7cm}
}
\caption[]{Uncorrected (left) and corrected(right) star ellipticities in
one of cosmic shear fields.
\label{F14stars.ps}}
\end{figure}

Unfortunately, the ellipticity of the galaxies measured on the
images are contaminated by atmospheric and instrumental
distortions of the Point Spread Function (PSF) that 
also produce coherent non-gravitational elongation
patterns, even on stars.
Such example of PSF is displayed on Figure \ref{psf.ps}, and the measured
coherence of the PSF distortion on a real field is shown on Figure 
 \ref{F14stars.ps}.
This is a critical issue, for instance
there were two early tentatives to measure the gravitational lensing by
large scale structures, which failed because the image quality was
very low and the PSF correction not accurate
(Valdes et al. 1983, Mould et al. 1994). Since then, various methods have been
developped to correct for the non gravitational source of galaxy alignment,
which followed the improvement of image quality:

\begin{itemize}
\item{} Kaiser et al. (1995), a method which treats the PSF convolution analytically
to the first order. It is called KSB.
\item{} Bonnet \& Mellier (1995), which combines galaxy image simulation and optimal
weighting of the isophotes.
\item{} The auto-correlation function (Van Waerbeke et al. 1997), similar to
Bonnet \& Mellier (1995), but applied to the auto-correlation of the image
of the galaxies
to avoid some problems associated with the galaxies.
\item{} Kuijken (1999), a method which parametrizes the PSF and the galaxies
with analytical functions, and try to match the convoled profile to the data.
\item{} Kaiser (2000), extended KSB, which circularises the PSF
before the isotropic correction.
\item{} Modified KSB (Rhodes et al. 2000), is the KSB approach, applied on the
galaxy moments instead of the ellipticities.
\item{} Bernstein \& Jarvis (2002), is first a circularisation of the PSF, and then
the convolved profile is analysed using a reduced set of orthogonal functions
(Laguerre polynomials).
\item{} The shapelets approach (Chang \& R\'efr\'egier, 2002), is a kind
of Principal Components Analysis, using orthogonal Hermite polynomials functions
to decompose the convolved galaxy images (see also Bertin 2001 for a PCA
approach).
\end{itemize}

\begin{figure}[t]
\centerline{
\psfig{figure=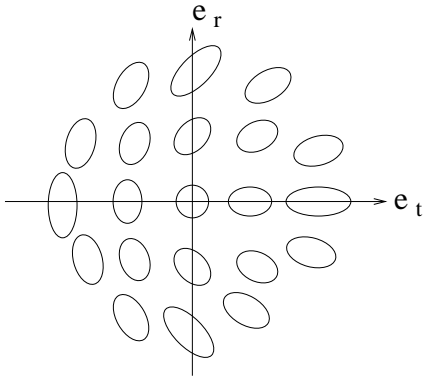,height=5cm}
}
\caption[]{Value of ($e_t$,$e_r$), or ($e_1$, $e_2$) in Cartesian coordinates,
as a function of the shape of a galaxy with respect to the local frame attached to the
galaxy. Note that the ellipticity is invariant by a rotation of $\pi$, and not
$2\pi$, this is why $e_t<0$ and $e_r=0$ for a vertical galaxy for instance.
\label{ellip.eps}}
\end{figure}

The most popular, and certainly the most intensively tested \footnote{A realistic
image simulation software is available at http://affix.iap.fr/soft/skymaker/index.html,
and a realistic catalogue generation at http://affix.iap.fr/soft/stuff/index.html}
(Erben et al. 2001,
Bacon et al. 2001), is the KSB approach.
It is a very simple and powerful correction based on the first order
effect of a convolution. The idea is that we can write the first order effect of the
shear and of the PSF convolution analytically as:

\begin{equation}
\eg^{obs}=\eg^{source}+{\rm P}_\gamma\cdot\gammag+{\rm P}^{sm}\cdot \eg^{\star},
\label{eobs}
\end{equation}
where ${\rm P}_\gamma$ and ${\rm P}^{sm}$ are tensors computed on the image
(see Kaiser et al., 1995),
$\eg^{\star}$ is the star ellipticity at the galaxy location, and $\gammag$ is the shear
signal we want to measure.
Assuming that the galaxies are isotropically oriented in the source plane,
we have $\langle\eg^{source}\rangle=0$ (which is valid even if the
galaxies are intrinsically correlated), therefore the shear estimate from the
measured galaxy ellipticity is given by:

\begin{equation}
\gammag={\rm P}_\gamma^{-1}\left(\eg^{obs}-{\rm P}^{sm}\cdot \eg^{\star}\right).
\end{equation}
We discussed in the previous section how the shear ($\gammag$) could be splitted
into a radial and a tangential component $\gamma_r$ and $\gamma_t$ when
projected onto the local frame of the aperture (Figure \ref{filter.eps}).
Figure \ref{ellip.eps} shows the relation between the components $\eg=(e_1,e_2)$
of a galaxy, and its orientation. If we identify $(e_1,e_2)$ to $(e_t,e_r)$,
we obtain the orientation in the local frame.


\subsubsection{E and B modes}

\begin{figure}[t]
\centerline{
\psfig{figure=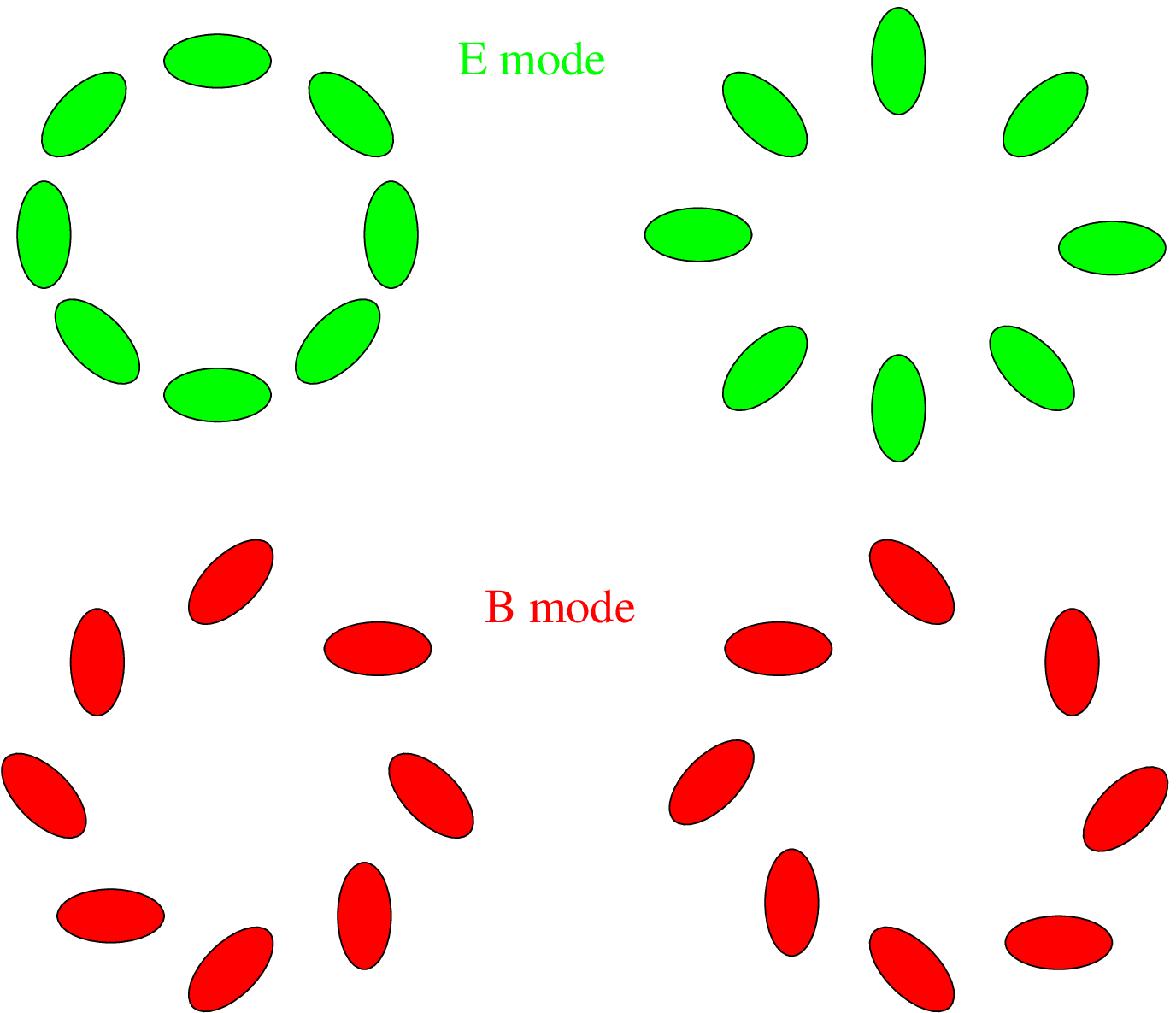,height=8cm}
}
\caption[]{Top patterns: shear curl free modes (E modes) allowed by gravitational
lensing. Bottom
patterns: curl modes (B modes) not allowed from a scalar gravitational potential.
Only the E modes gives the signal of
the aperture mass statistics  $\langle M_{\rm ap}^2\rangle$.
\label{EBmode.eps}}
\end{figure}

The gravitational field is supposed to be completely dominated by a scalar 
   gravitational
potential at low redshift. The consequence is that only curl free modes for 
   the
shear are allowed. Any significant curl component should be interpreted as
a (bad) sign of residual systematics in the data. Figure \ref{EBmode.eps} shows
the E mode generated by over-densities (top-left) and under-densities (top-right).
The two bottom curl modes are not allowed.
Using the statistical properties of these patterns and the 
    ($e_t$,$e_r$) conversion
from Figure \ref{ellip.eps}, it can be shown that the E modes
correspond  to the aperture mass $\langle M_{\rm ap}^2\rangle$, 
   and the B mode
to the aperture mass with the galaxies $45$ degrees rotated (such rotation 
  corresponds
to a switch $e_t\rightarrow e_r$; $e_r\rightarrow -e_t$). This is easy to understand:
if there is no B mode, then switching the E into B, and B into E modes kills
the signal measured with the aperture mass statistics.

\subsubsection{Aperture mass from the shear correlation function}

\begin{figure}[t]
\centerline{
\psfig{figure=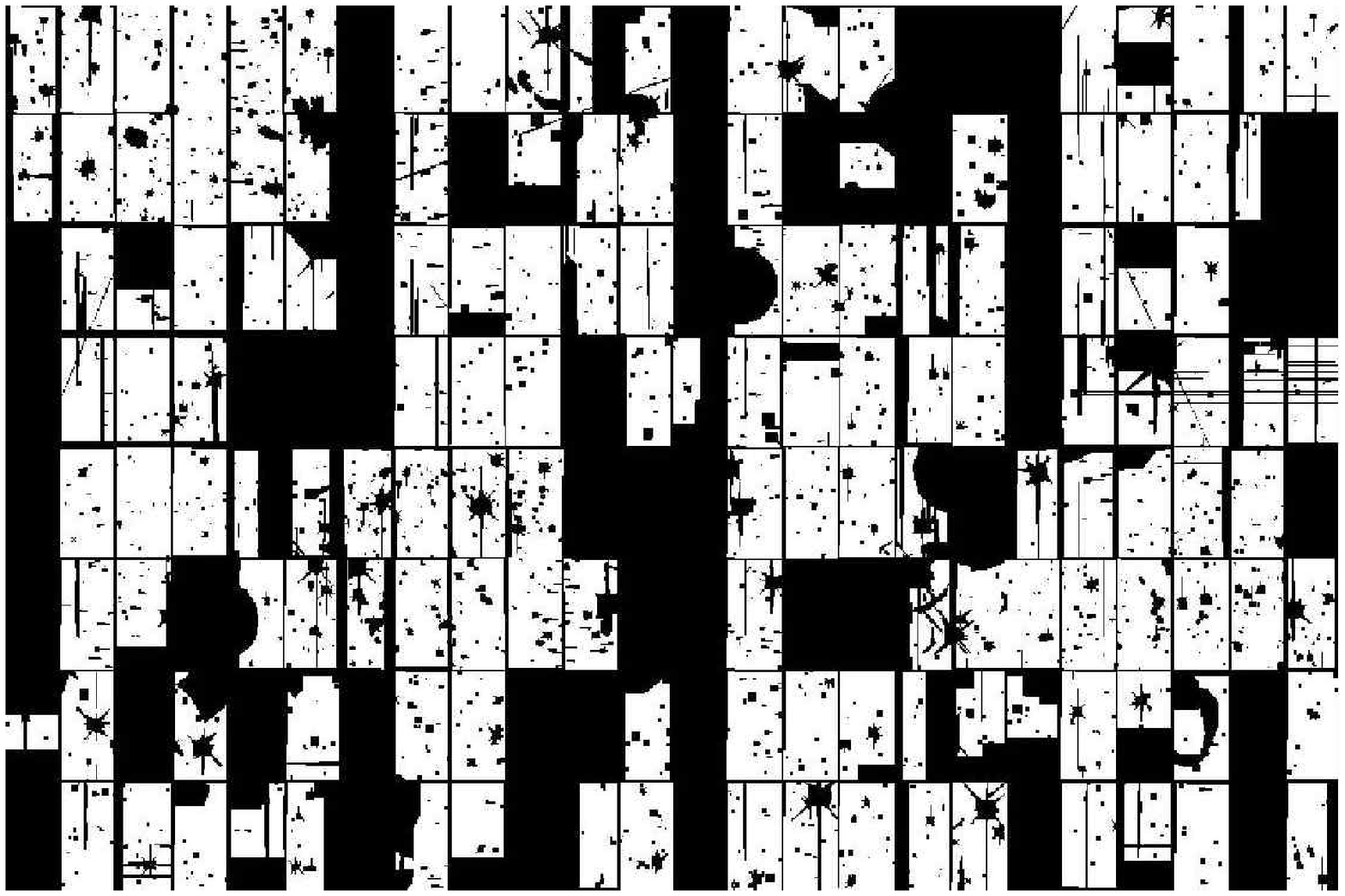,height=8cm}
}
\caption[]{Mask area for real data. These holes of various sizes make the mass
reconstruction very challenging. Each CCD chip is about $7'\times 14'$. Entire CCD's
had to be removed because of bright stars and residual fringes patterns.
\label{maskmap.ps}}
\end{figure}

Because the E/B mode separation provides a direct and robust 
  check of systematics error residuals,
 it is widely believed to be the most 
  reliable statistics. In order to compute it, there
is fortunately no need to draw a compensated filter across the data and to 
   average
the shear variance; otherwise, this could be terribly complicated with 
    real data because
of the complex shape of the masks (see Figure \ref{maskmap.ps}). Variances and
correlation functions can be expressed one
into another (since they are only linear combinaison one to another).  
The $E$ mode aperture mass is given by

\begin{equation}
\langle M_{\rm ap}^2\rangle=\pi\int_0^{2\theta_c} rdr {\cal W}(r) \xi_+(r)+
\pi\int_0^{2\theta_c} rdr \tilde{\cal W}(r) \xi_-(r),
\label{MapE}
\end{equation}
where ${\cal W}(r)$ and $\tilde{\cal W}(r)$ are given in Crittenden et al. 2002
and Pen et al. 2002.
The $B$-mode
is obtained by changing the sign of the second term in Eq.(\ref{MapE}). The
correlation functions $\xi_+(r)$ and $\xi_-(r)$ are compute from the tangential
and radial correlation functions (see Eq.\ref{corrfuncdef}).
In order to estimate the shear correlation functions, let $\thetag_i$
be location of the $i$-th galaxy, its
ellipticity $\eg(\thetag_i)=(e_1,e_2)$, and the weight $w_i$.  The
ellipticity is an unbiased estimate of the shear $\gammag(\thetag_i)$.
The quantity measured from the data are the binned tangential and
radial shear correlation functions.  They are given by a sum over
galaxy pairs $(\thetag_i,\thetag_j)$

\begin{equation}
\xi_{tt}(r)={\displaystyle\sum_{i,j} w_i w_j e_t(\thetag_i)\cdot e_t(\thetag_j)
\over \displaystyle\sum_{i,j} w_i w_j} \ \ \ ; \
\xi_{rr}(r)={\displaystyle\sum_{i,j} w_i w_j e_r(\thetag_i)\cdot e_r(\thetag_j)
\over \displaystyle\sum_{i,j} w_i w_j},
\label{corrfctestim}
\end{equation}
where $r=|\thetag_i-\thetag_j|$, and $(e_t, e_r)$ are the tangential and radial
ellipticities defined in the frame of the line connecting a pair of galaxies.  
The weights $w_i$ are usually computed for each galaxy from the intrinsic 
   ellipticity
variance $\sigma_e^2$ and the r.m.s. of the ellipticity PSF correction
$\sigma_\epsilon^2$. For example, van Waerbeke et al (2000) 
   measured
   $\sigma_e\simeq 0.4$ from their CFHT data, and
 defined the weights as:

\begin{equation}
w_i={1\over \sigma_e^2+\sigma_\epsilon^2}.
\end{equation}
To compute $\sigma_\epsilon$ for each galaxy, the galaxy
size-magnitude parameter space is divided into cells of constant object number
(typically $30$ galaxies per cell). For each cell the
r.m.s. of the ellipticity correction among the galaxies in the cell
is computed.
This choice of parameter space is motivated by the fact that the
isotropic PSF correction (the $P_\gamma$ term in Eq.\ref{eobs}) is mainly
sensitive to the size and magnitude of the galaxies.

\section{2-pts statistics}

\subsection{Measurements}

\begin{figure}[t]
\centerline{\vbox{
\psfig{figure=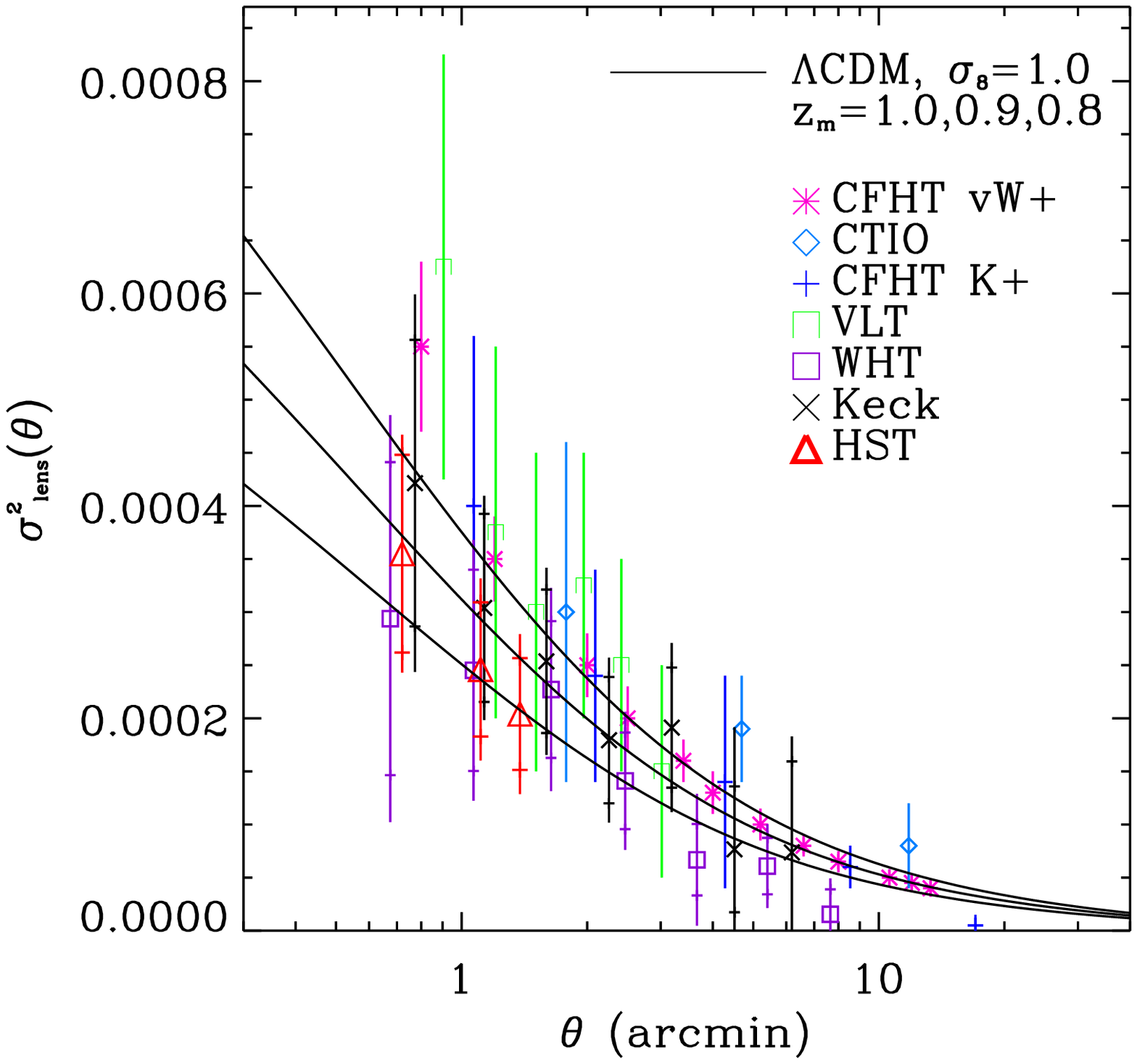,height=8.cm}
}}
\caption[]{Compilation of recent results of top-hat shear variance
measurements from several groups (Courtesy R\'efr\'egier et al. 2002).
\label{ref02.eps}}
\end{figure}
\begin{figure}[t]
\centerline{
\psfig{figure=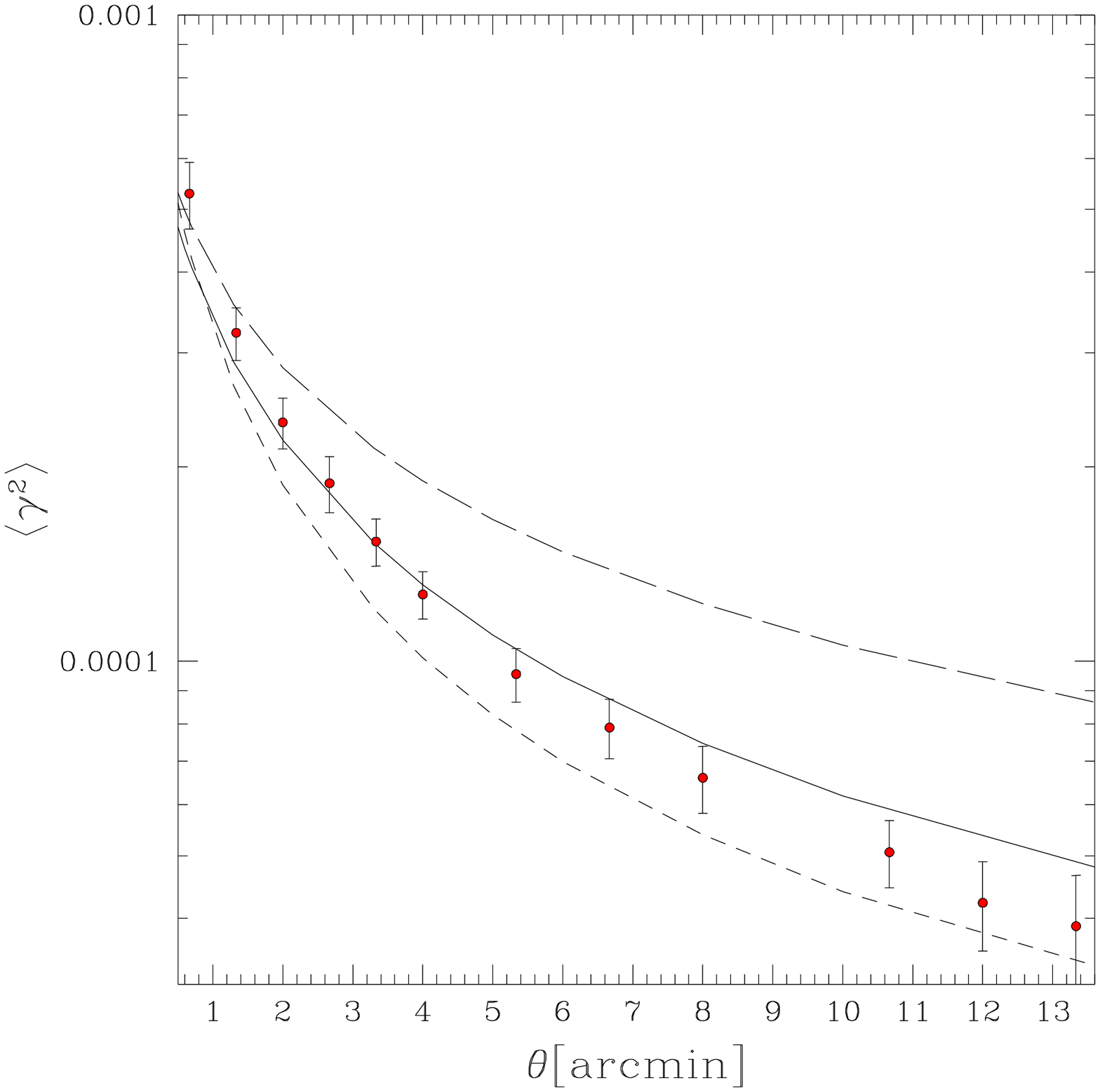,height=6.5cm}
\psfig{figure=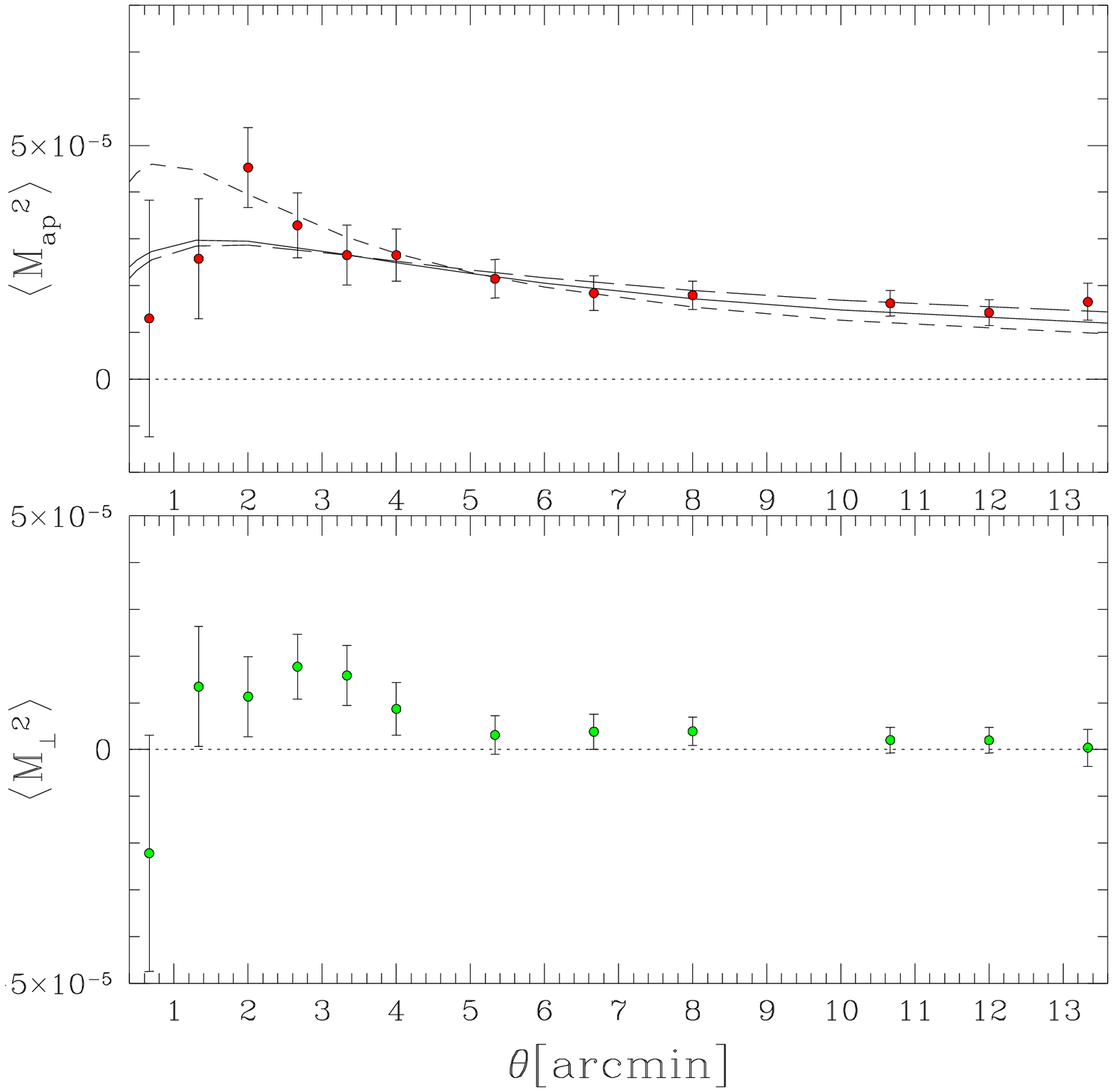,height=6.5cm}}
\centerline{
\psfig{figure=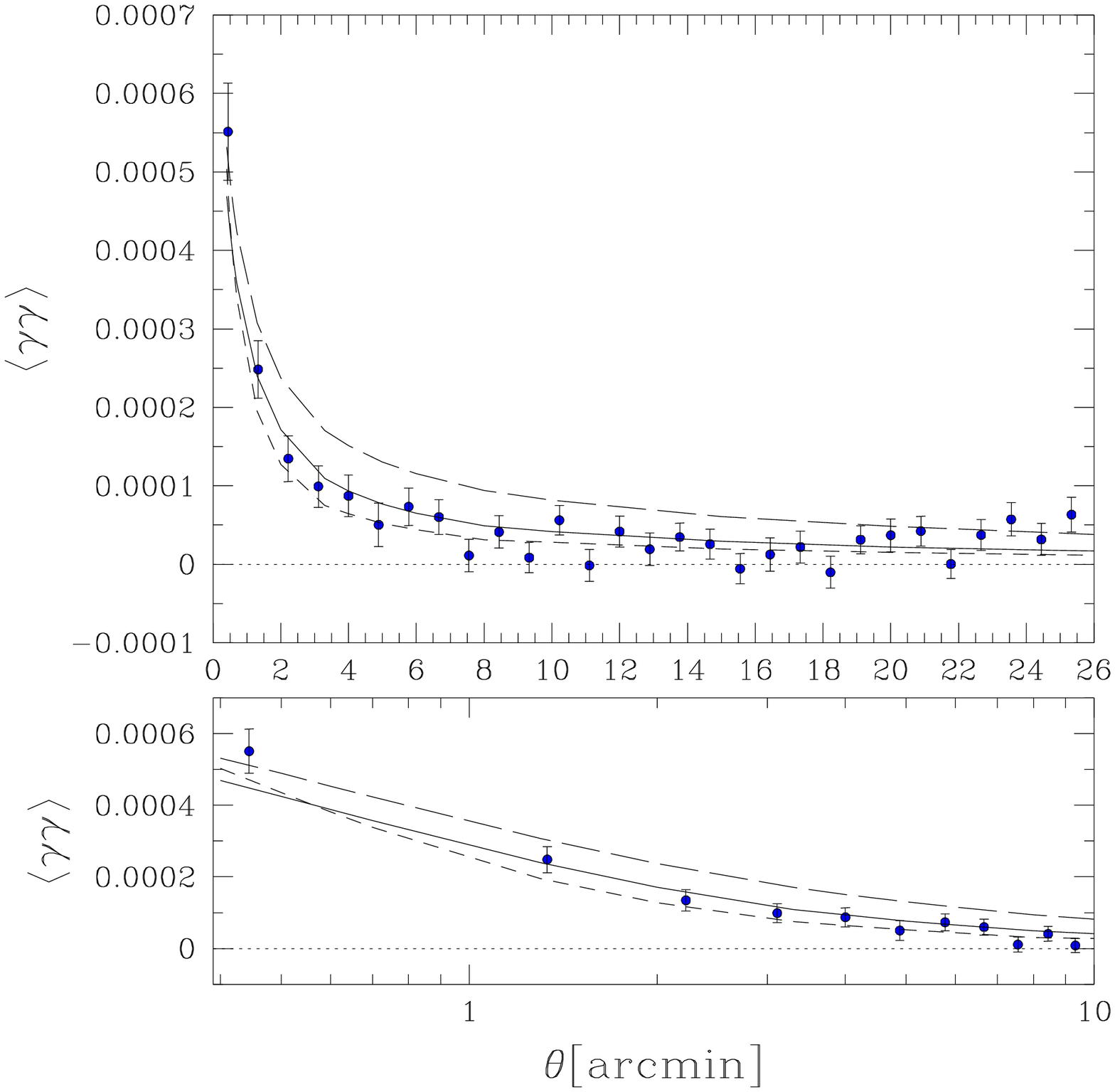,height=6.5cm}
\psfig{figure=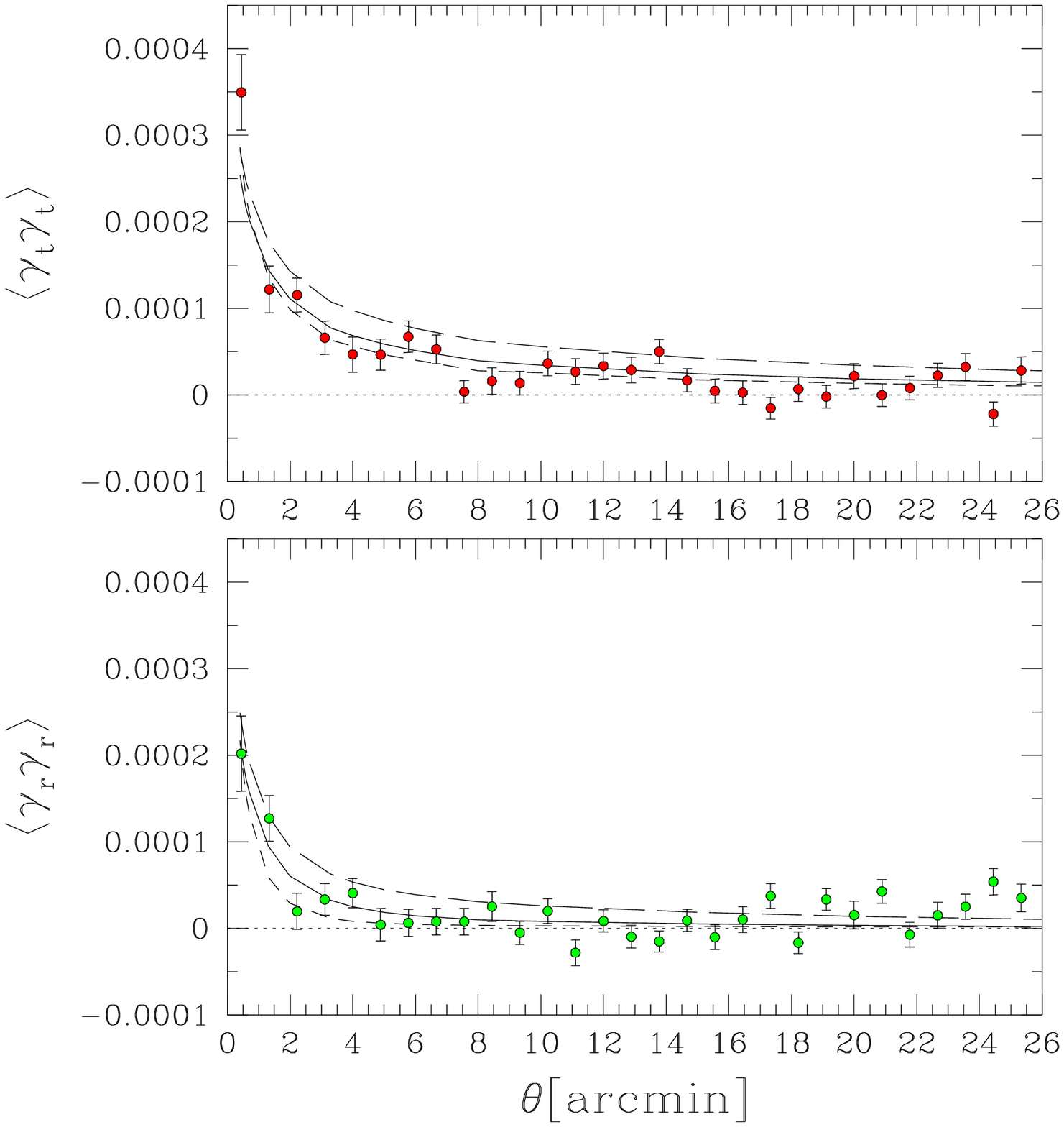,height=6.5cm}}
\caption{Measurement of all the 2-points statistics in the same
survey, VIRMOS (Van Waerbeke et al. 2001).
Top left: top-hat variance. Top right: aperture
mass $E$ and $B$ modes. Bottom left: full shear correlation
function. Bottom right: projected shear correlation functions.
survey. Right: $E$ (top) and $B$ (bottom) modes measured in the
RCS survey. The $B$ mode is low and the $E$ mode compatible
with the predictions for the aperture mass statistics. The lines
are fiducial models which indicate the relative deviations
between the statistics to the models.
\label{self.ps}}
\end{figure}
\begin{figure}[t]
\centerline{
\psfig{figure=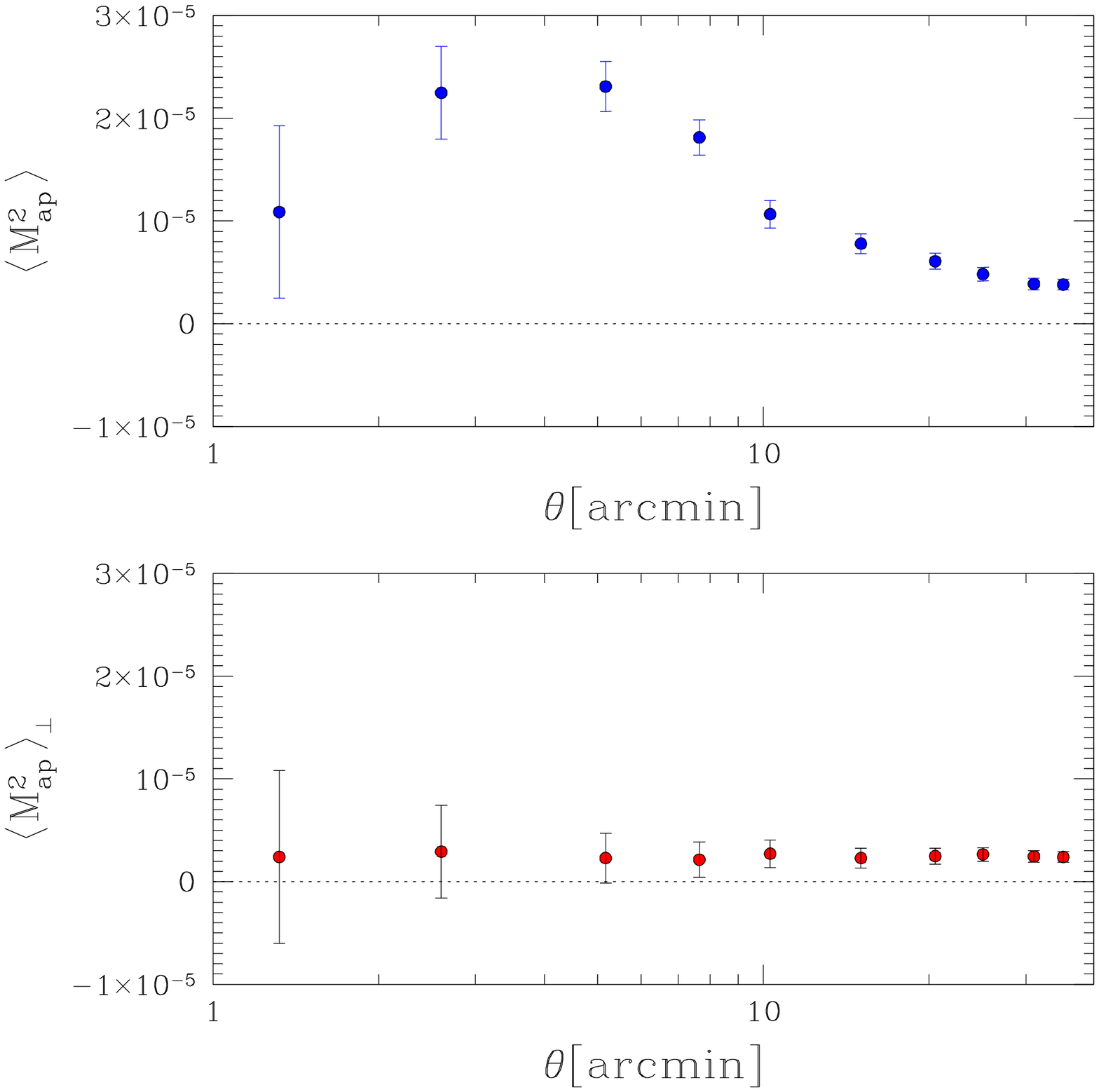,height=6cm}
\psfig{figure=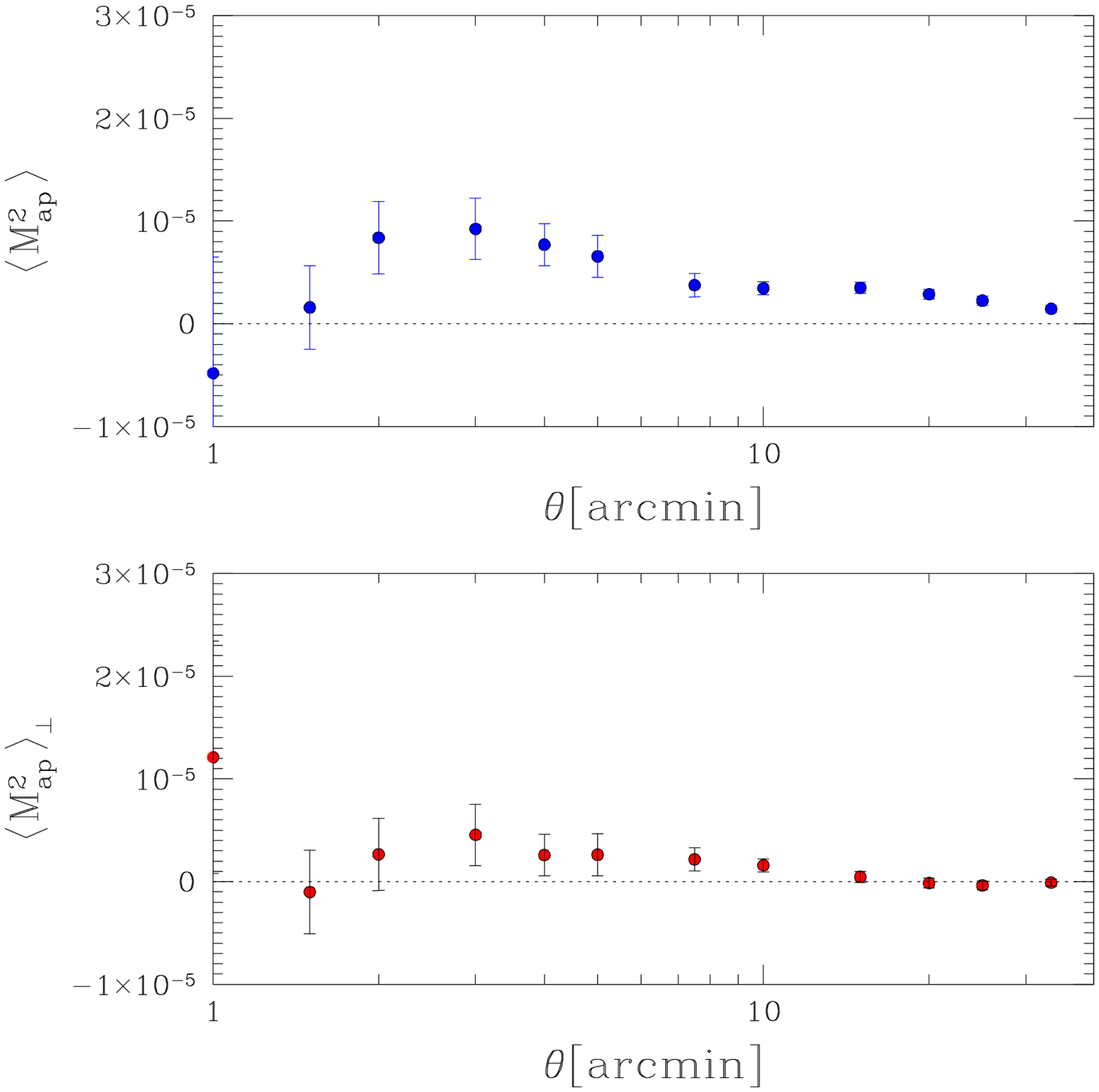,height=6cm}}
\caption{Left: $E$ (top) and $B$ (bottom) modes measured
with all the most recently reduced data in the VIRMOS
survey (Van Waerbeke et al. 2002). Right: $E$ (top) and $B$ (bottom) modes measured in the
RCS survey (Hoekstra et al. 2002). The $B$ mode is low and the $E$ mode compatible
with the predictions for the aperture mass statistics.
\label{mapplot}}
\end{figure}

There are
now several evidences of the cosmological origin of the measured signal:

(a) The consistency of the shear excess variance measured from different telescopes,
at different depths and with different filters. This is summarized on Figure
\ref{ref02.eps}. The first detections were obtained by Bacon et al. 2000,
Kaiser et al. 2000, Van Waerbeke et al. 2000, Wittman et al. 2000.
Since then, several measurements have been done in different observing conditions,
which are summarized in Table \ref{constraints}.

(b) On a single survey, the self consistency of the different types of lensing
statistics given by
Eqs.(\ref{tophatstat},\ref{mapstat},\ref{ggstat},\ref{etetererstat}). This was
done on the VIRMOS-DESCART survey \footnote{http://www.astrsp-mrs.fr and
http://terapix.iap.fr/DESCART}, and it is 
shown in Figure \ref{self.ps} (Van Waerbeke et al. 2001).

(c) The comparison of the $E$ and $B$ modes measurements
(to higher accuracy
than in (b)) between a deep and shallow survey
for the VIRMOS-DESCART and RCS
\footnote{http://www.astro.utoronto.ca/~gladders/RCS/} surveys
(Van Waerbeke et al. 2002, Hoekstra et al. 2002). This is shown on
Figure \ref{mapplot}.
More recently, the $E$ and $B$ modes have been also measured in other surveys
(Brown et al. 2003, Jarvis et al. 2003, Hamana et al. 2003), which supports
the cosmological origin of the
signal, showing also the already small amount of residual systematics
achieved with today's technology. The $E$ and $B$ mode measurements should now
be considered as the most robust proof of the cosmological origin of the signal,
and a quantitative test of systematics.

(d) The lensing signal is expected to decrease for low redshift sources,
as consequence of the lower efficiency of the the gravitational
distortion. It corresponds to a change in $w_s$ in Eq(\ref{kappaeq}),
or equivalently a change in the mean source redshift with Eq(\ref{pofkappa}).
This decrease
of the signal has been observed for the first time with the comparison
of the $E$ mode amplitude of the
VIRMOS survey aperture mass (see Figure \ref{mapplot}), which has a source mean redshift
around $0.9$, to the RCS which has a source mean redshift around $0.6$. The
expected decrease in signal amplitude is about $2$, which is what is observed.
This is a direct evidence of the effect of changing the redshift of the sources,
a kind of 3-D cosmic shear effect.

(e) Space images provide in principle a systematics-low environment, and even if
the observed areas are still smaller than ground based observations, space
data provide ideal calibrations of the cosmic shear signal
(Rhodes et al. 2001, Haemmerle et al. 2002, R\'efr\'egier et al. 2002),
which are in agreement with ground based measurements
(see Figure \ref{ref02.eps},
the HST points).

\subsection{Constraints}

The standard approach is to compute the likelihood of a set of $n$ parameters
$(p_1,p_2,...,p_n)$, knowing the data vector $\dg$, as written in
Eq(\ref{likelihood}). As the data vector, it is natural to choose the aperture
mass variance as a function of scale $\langle M_{\rm ap}^2\rangle$, because
the signal is splitted into gravitational lensing and systematics channels (the
$E$ and $B$ modes). The $B$ mode measures an estimate of the contamination
of the $E$ mode by systematics.  The $E$ and $B$ modes do not 
  equally contribute to systematic, but we know, from the measurement
of the modes on the stars, that they are very similar. If the $B$ mode is not
consistent with zero (which is the case for all surveys at the moment),
it is important to deal with it properly when estimating the
cosmological parameters. Unfortunately it is not yet clear what the best approach
is: some groups (Van Waerbeke et al. 2002, Hoekstra et al. 2002, 
   Hamana et al. 2003) added the $B$ mode in
quadrature to the $E$ errors, taking into account the correlation between
various scales. The $B$ mode has been subtracted first from the $E$ mode in
Hoekstra et al. (2002), but not in Van Waerbeke et al. (2002). This might 
   probably result in a slight
bias for high $\sigma_8$ values in the later. Unfortunately we have no guarantee
that the $B$ subtraction is the right correction method.
Recently Jarvis et al. (2003) marginalised the probabilities
over $E-B$ to $E+B$ taken as the signal, which is more likely to include
the 'true' $B$ mode correction one has to apply.

\begin{figure}
\centerline{\vbox{
\psfig{figure=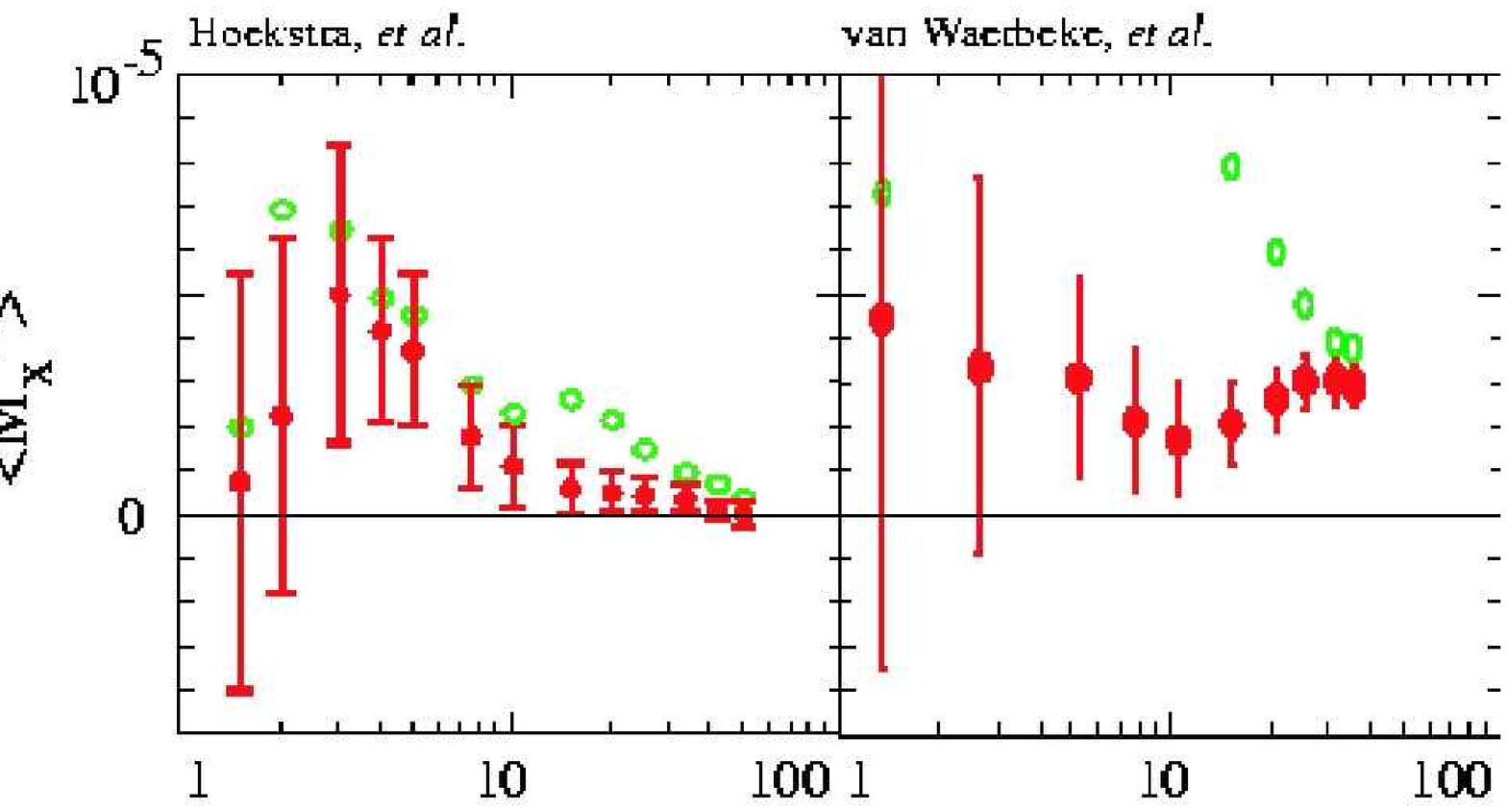,height=6.cm}
\psfig{figure=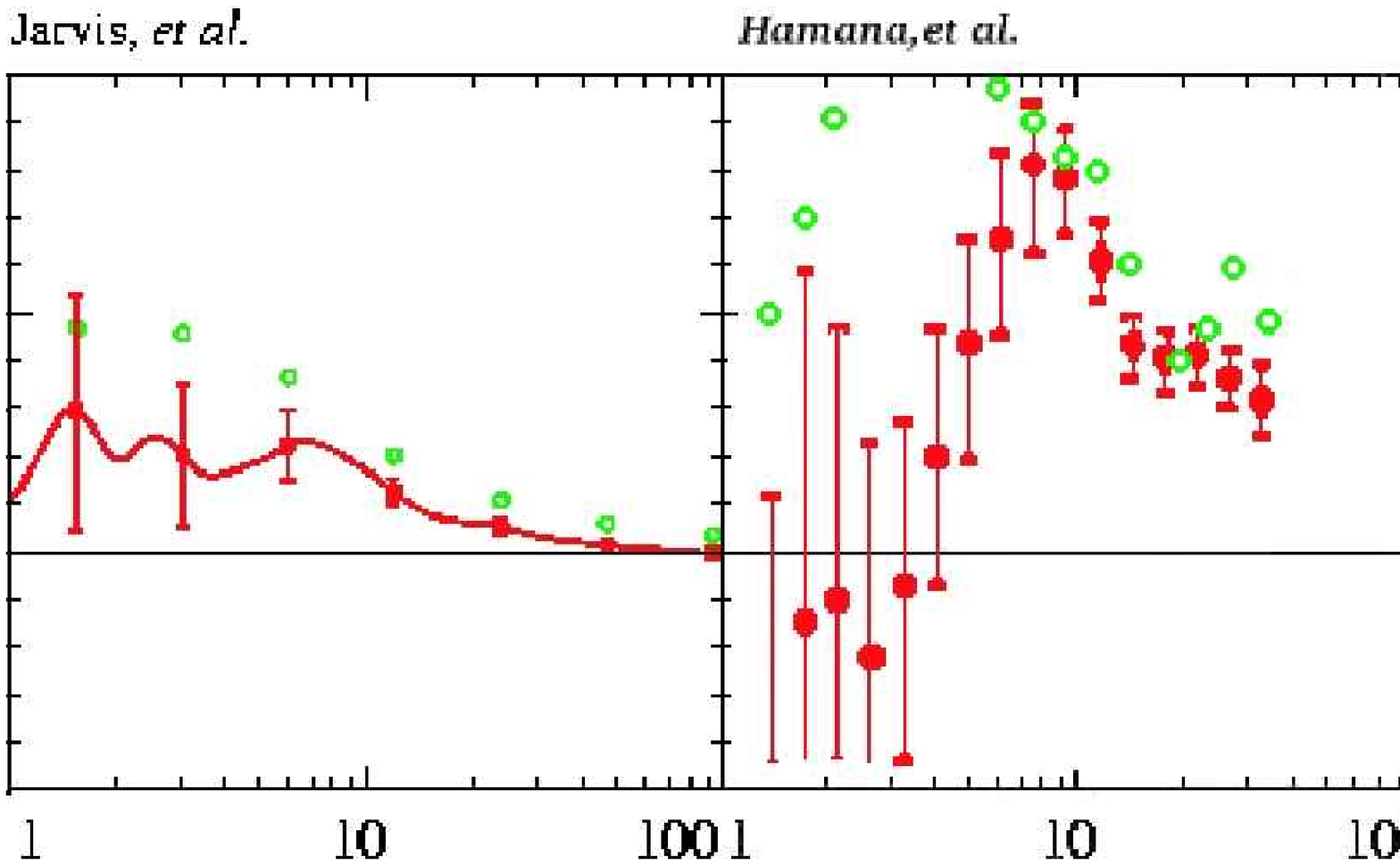,height=6.cm}
}}
\caption[]{Plot showing the relative amplitude of the aperture mass
$E$ and $B$ modes (points without and with error bars respectively) for all
the surveys where
the aperture mass has been measured (Hoekstra
et al. 2002, Van Waerbeke et al. 2002, Jarvis et al. 2003
and Hamana et al. 2003)
(picture taken from Jarvis et al. 2003 and extended). The result of
Hoekstra et al. (2002) is for the full magnitude range, while in Figure
\ref{mapplot}, right panel, it is for the galaxies used for the cosmic shear analysis.
\label{bmode4ex.ps}}
\end{figure}
Figure \ref{bmode4ex.ps} shows
\footnote{The $B$ mode peak
at $10'$ in Hamana et al. (2003) is due to a PSF correction error over the mosaic.
It is gone when the proper correction is applied, Hamana, {\it private communication}.}
the $E$ and $B$ modes
that have been measured so far, using the aperture mass only (this is
the only statistic which provides an unambiguous $E$ and $B$ separation,
Pen et al. 2002). The two deepest
surveys have large scale
$B$ mode contamination (Van Waerbeke et al. 2002, Hamana et al. 2003), and the two shallow surveys have
small scale contamination (Hoekstra et al. 2002, Jarvis et al. 2003).

\begin{figure}[t]
\centerline{
\psfig{figure=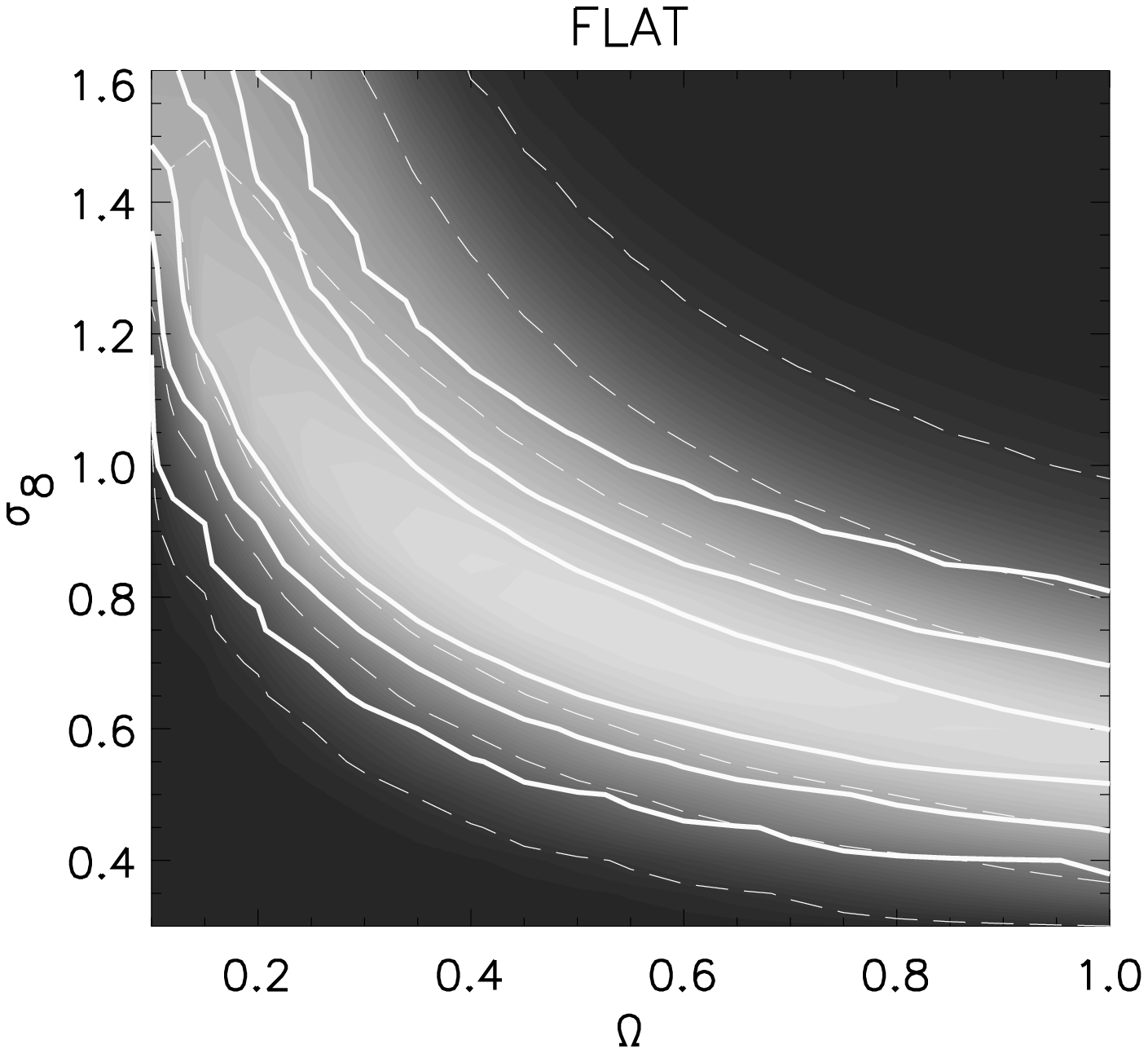,height=5.cm}
\psfig{figure=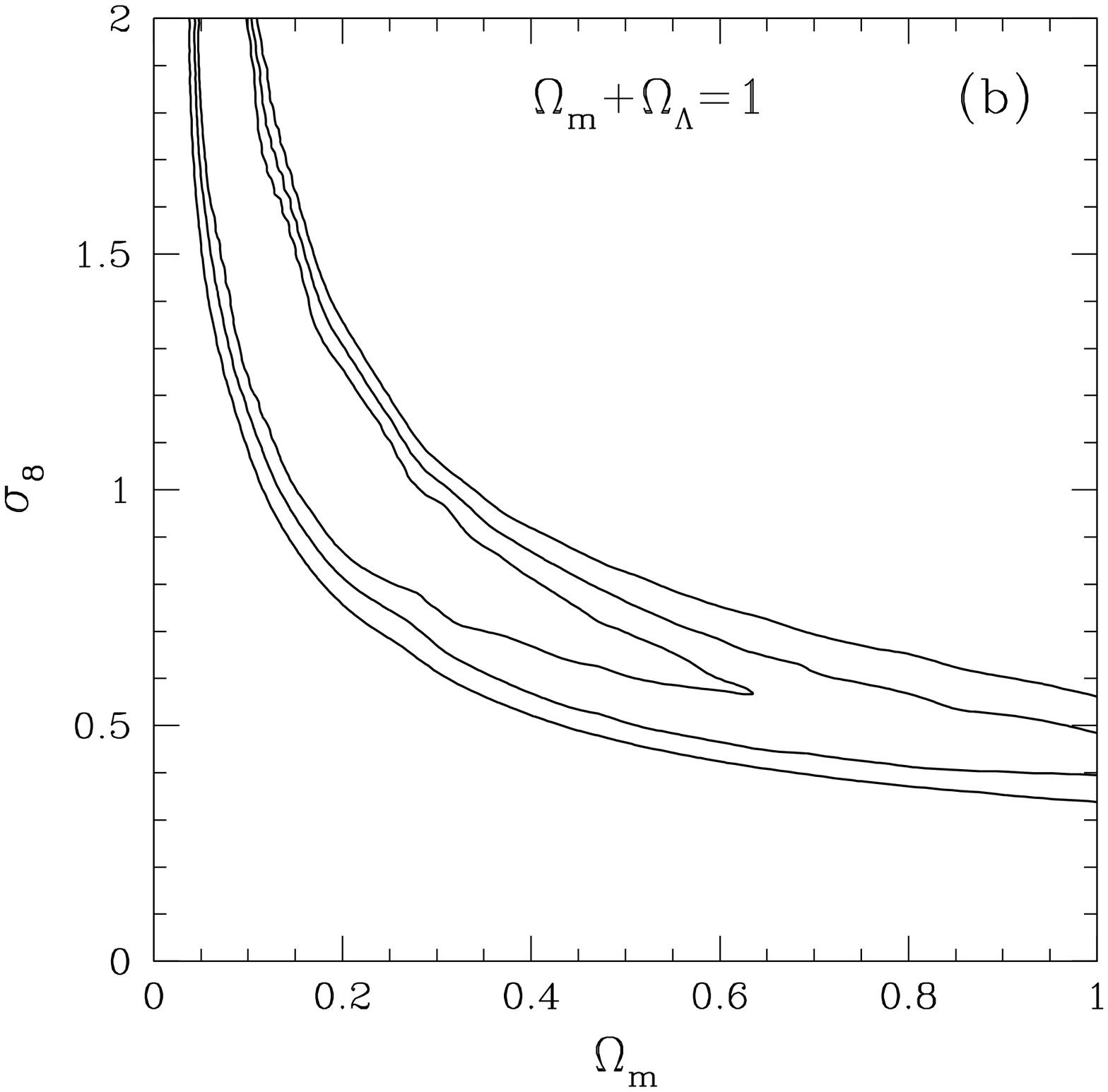,height=5.cm}
}
\caption[]{The solid lines on each plot show the 1, 2 and 3$\sigma$
contours of the VIRMOS and RCS survey, from the measurements shown in
Figure \ref{mapplot}. The contours have been marginalised over the source
redshift and the slope of the matter power spectrum as described elsewhere
(Van Waerbeke et al. 2002, Hoekstra et al. 2002).
\label{flat_omega_sigma8.ps}}
\end{figure}

Figure \ref{flat_omega_sigma8.ps} shows the joint
$\Omega_m$, $\sigma_8$ constraints
obtained from the measurements of Figure \ref{mapplot}. They are
obtained only when comparing the measured lensing signal to the non-linear
predictions. Unfortunately, the actual surveys are not yet big enough to probe the
linear scales accurately. The non-linear power can be computed numerically
(Smith et al. 2002), but its precision is still uncertain. Recent investigations
show that a $10 \%$ r.m.s. uncertainty is expected, which means that the
cosmological parameters cannot be known with better precision for the moment.
According to the Figure \ref{stats.ps}, the transition scale
between the linear and non-linear regimes is around 1 degree. The consequence
is that
the quoted mass normalization $\sigma_8$ is sensitive to the validity
of the non-linear mapping at small scale. In this respect, Jarvis et al. (2003)
are less contaminated by this problem because they used the lensing signal
from $30'$ to $100'$ to constrain the mass normalization.

Table \ref{constraints} summarizes
the $\sigma_8$ measurements for all the lensing surveys published so far. For
simplicity it is given for $\Omega_m=0.3$. Despite the differences among the
surveys, it is worth to note that the results are all consistent within
$2.5\sigma$ between the most extreme cases, when poorly known parameters are
marginalised.

\begin{landscape}
\begin{table*}[t]
\begin{center}
{\small
\caption{Constraints on the power spectrum normalization "$\sigma_8$" for
$\Omega_m=0.3$ for a flat Universe, obtained from a given "statistic".
"CosVar" tells us whether or not the cosmic variance
has been included, "E/B" tells us whether or not a mode decomposition has been used
in the likelihood analysis. Note that Van Waerbeke et al. (2001) and Brown et al. (2003)
measured a
small B-mode, which they didn't use in the parameter estimation. $z_s$ and
$\Gamma$ are the priors used for the different
surveys identified with "ID". Note also the cosmic shear results obtained by
Kaiser et al. (2000) and Haemmerle et al. (2002), which
are not in the table here because they reported a shear detection, not a $\sigma_8$
measurement.
\label{constraints}
}
\label{tabcs}
\bigskip
\begin{tabular}{lcccccccc}\hline
\\
ID & $\sigma_8 $ & Statistic & Field & $m_{\rm lim}$& CosVar & E/B & $z_s$ & $\Gamma$ \\
\hline
Maoli  et al. 01 & $1.03\pm 0.05$ & $\langle\gamma^2\rangle$ & VLT+CTIO & - & no & no & - & 0.21 \\
 &  &  & +WHT+CFHT & & & & & \\
\hline
Van Waerbeke et al. 01& $0.88\pm 0.11$ & $\langle\gamma^2\rangle$, $\xi(r)$, $\langle M_{\rm ap}^2\rangle$  & CFHT 8 sq.deg.& I=24 & no & no (yes) & 1.1 & 0.21 \\
\hline
Rhodes  et al. 01& $0.91^{+0.25}_{-0.29}$ & $\xi(r)$ & HST 0.05 sq.deg.& I=26 & yes & no & 0.9-1.1 & 0.25 \\
\hline
Hoekstra et al. 02& $0.81\pm 0.08$ & $\langle\gamma^2\rangle$ & CFHT+CTIO & R=24 & yes & no & 0.55 & 0.21 \\
 &  & & 24 sq.deg. &  & &  &  &  \\
\hline
Bacon et al. 03& $0.97\pm 0.13$ & $\xi(r)$ & Keck+WHT & R=25 & yes & no & 0.7-0.9 & 0.21 \\
 &  &  & 1.6 sq.deg. & & & & & \\
\hline
R\'efr\'egier et al. 02 & $0.94\pm 0.17$ & $\langle\gamma^2\rangle$ & HST 0.36 sq.deg. & I=23.5 & yes & no & 0.8-1.0 & 0.21 \\
\hline
Van Waerbeke et al. 02& $0.94\pm 0.12$ & $\langle M_{\rm ap}^2\rangle$ & CFHT & I=24 & yes & yes & 0.78-1.08 & 0.1-0.4 \\
 &  &  & 12 sq.deg. &  & &  &   &   \\
\hline
Hoekstra et al. 02& $0.91^{+0.05}_{-0.12}$ & $\langle\gamma^2\rangle$, $\xi(r)$ & CFHT+CTIO & R=24 & yes & yes & 0.54-0.66 & 0.05-0.5 \\
 &  & $\langle M_{\rm ap}^2\rangle $ & 53 sq.deg. &  & &  &  &  \\
\hline
Brown et al. 03&  $0.74\pm 0.09$ & $\langle\gamma^2\rangle$, $\xi(r)$  & ESO 1.25 sq.deg.& R=25.5 & yes & no (yes) & 0.8-0.9 & - \\
\hline
Hamana et al. 03 & $(2\sigma) 0.69^{+0.35}_{-0.25}$ & $\langle M_{\rm ap}^2\rangle$, $\xi(r)$  & Subaru 2.1 sq.deg. & R=26 & yes & yes & 0.8-1.4 & 0.1-0.4 \\
\hline
Jarvis et al. 03 & $(2\sigma) 0.71^{+0.12}_{-0.16}$ & $\langle\gamma^2\rangle$, $\xi(r)$, $\langle M_{\rm ap}^2\rangle$  & CTIO 75 sq.deg. & R=23 & yes & yes & 0.66 & 0.15-0.5 \\
\hline
\end{tabular}
}
\end{center}
\end{table*}
\end{landscape}

\section{3-pts statistics}

\begin{figure}[t]
\centerline{\vbox{
\psfig{figure=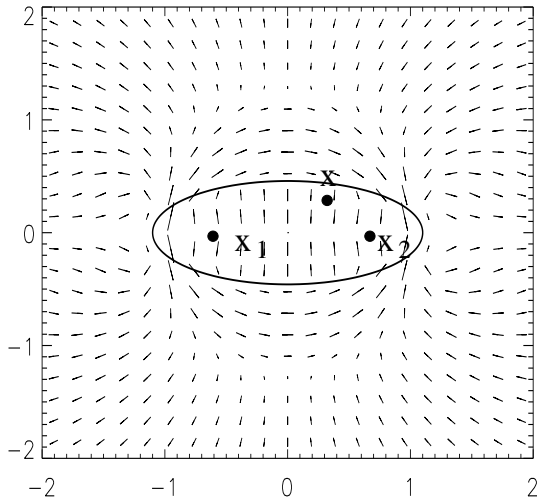,height=8.cm}
}}
\caption[]{Average shear pattern obtained around a galaxy pair located at
($\xg_1$, $\xg_2$). A third galaxy is located at $\xg$, its shear vector is
projected along the vertical axis, it is called $\gamma_t(\xg)$. The
shear 3-points function
$\langle \gammag(\xg_1)\cdot\gammag(\xg_2)\gamma_t(\xg)\rangle$
is averaged inside the ellipse indicated by the solid line.
\label{pattern.eps}}
\end{figure}
So far, we only discussed the 2-points statistics, but 
 recently  higher order statistics have been also developed
for cosmic shear (Bernardeau et al. 1997, Jain \& Seljak 1997).
If we were able to reconstruct the convergence from the shear (ellipticity) measured on
the galaxies, one could for instance measure the top-hat smoothed higher order
statistic easily. For instance, the skewness of the convergence, which is defined as

\begin{equation}
S_3(\kappa)={\langle \kappa^3\rangle\over \langle \kappa^2\rangle^2},
\end{equation}
is of great interest because this 
   suited ratio of moments makes this statistic nearly independent of the
 normalization and shape of the power spectrum  
(Bernardeau et al. 1997). A pedagogical way to compare the second and 
   third moments is
to compute $\langle\kappa^2\rangle$ and $S_3(\kappa)$ in the perturbation 
   theory, and with
a power law power spectrum. In that case, one finds

\begin{eqnarray}
\sigma_{\kappa}&\approx& 0.01\ \sigma_8\ \Omega_0^{0.8}\ 
\left({\theta_0\over 1{\rm deg.}}\right)^{-(n+2)/2}\ z_s^{0.75},\label{var_def}\\
s_3&\sim& {\langle\kappa^3\rangle\over \langle\kappa^2\rangle^2}\approx
40\ \Omega_0^{-0.8}\ z_s^{-1.35}.
\label{skew_def}
\end{eqnarray}
These are only approximated relations, which are not valid in the real (non-linear)
world, but it shows
that the skewness provides a direct geometrical test (dependence on $\Omega_0$), as long
as we know the redshift of the sources $z_s$. Combined with the second order moment, the
degeneracy between the power spectrum normalization and the density parameter can be broken with
the cosmic shear alone.

The skewness can also be predicted in the non-linear regime, as for the
2-points statistics, using a non-linear extension of the bispectrum (Scoccimarro
et al. 2002, Van Waerbeke et al. 2002). The problem with the skewness of the
convergence is that it cannot be measured on the data directly, and
one needs to reconstruct $\kappa$ from the shear first. This process is
unfortunately  sensitive to the survey geometry because the projected
mass reconstruction is essentially a non-linear process. Given the typical observed
field geometry, as shown on Figure \ref{maskmap.ps}, it is yet impossible
to perform a mass reconstruction with the accuracy required to measure the
cosmic shear effect. One possibility for avoiding the mass reconstruction (that
is try to make the map making a local process) is to compute the third
moment of the aperture
mass (Schneider et al. 1998). Unfortunately, in that case as well, the complicated
survey geometry make it difficult to measure an accurate third moment
$\langle M_{\rm ap}^3\rangle$.

The alternative is to measure a third moment on the shear field
itself, but this cannot
be done in a trivial way, since for evident symmetry reasons, any odd moment
of the components of a vector field vanishes. One has to built explicitly
non-trivial measures of the third moment of the shear, which has been
recently proposed. So far, two of the proposed estimators lead to a measurement
(Bernardeau et al. 2002 and Pen et al. 2002).

In Bernardeau et al. (2002) the idea is to identify regular shear patterns around
any pair of lensed galaxies. A pair is identified by the two galaxy positions
$\xg_1$ and $\xg_2$, and any location around the pair by $\xg$. For a fixed
pair ($\xg_1$, $\xg_2$), we are interested in the average shear at $\xg$.

Figure \ref{pattern.eps} shows the typical shear pattern observed
around a galaxy pair located at ($\xg_1$, $\xg_2$). Ray tracing simulations
demonstrate the stability of this shear pattern, which is almost independent
on the cosmological model and the pair separation. A natural 3-points
function to calculate is the average of the product of the shear correlation
function $\gammag(\xg_1)\cdot\gammag(\xg_2)$ with a projection of the shear
of the third galaxy $\gamma(\xg)$. It is obvious from Figure \ref{pattern.eps}
that the projection is optimal when performed along the vertical axis. For a fixed
pair location ($\xg_1$, $\xg_2$), the 3-points function $\xi_3(\xg)$ is defined as:

\begin{equation}
\xi_3(\xg)=\langle \gammag(\xg_1)\cdot\gammag(\xg_2)\gamma_t(\xg)\rangle,
\end{equation}
and the quantity we measure is:

\begin{equation}
\overline{\xi_3}(\vert\xg_1-\xg_2\vert)= \int_{\rm
Ell.}{{\rm d}^2\xg'\over V_{\rm Ell.}} \xi_3(\xg').
\label{xi3_ellip}
\end{equation}

\begin{figure}
\centerline{
\psfig{figure=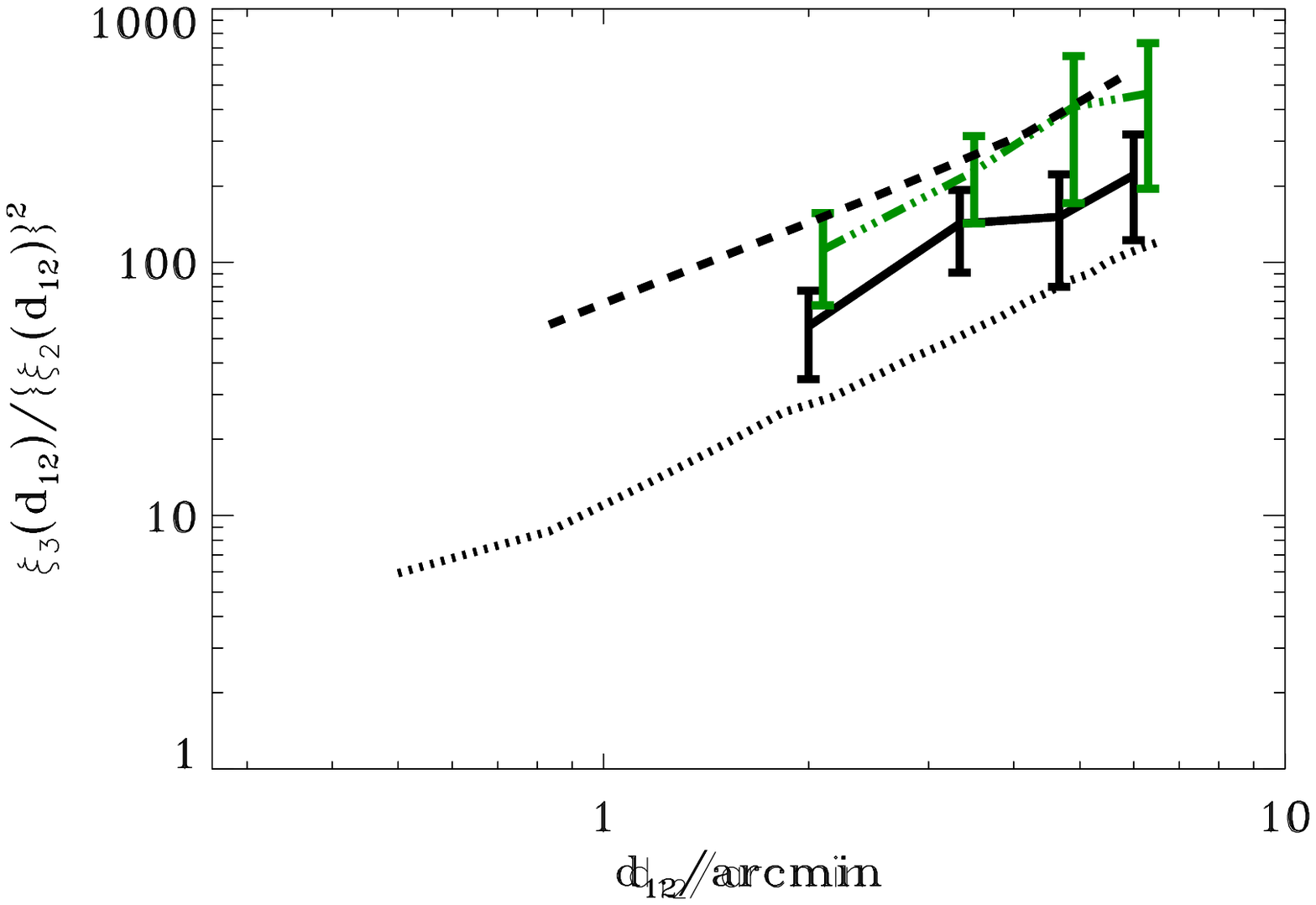,height=5.cm}
\psfig{figure=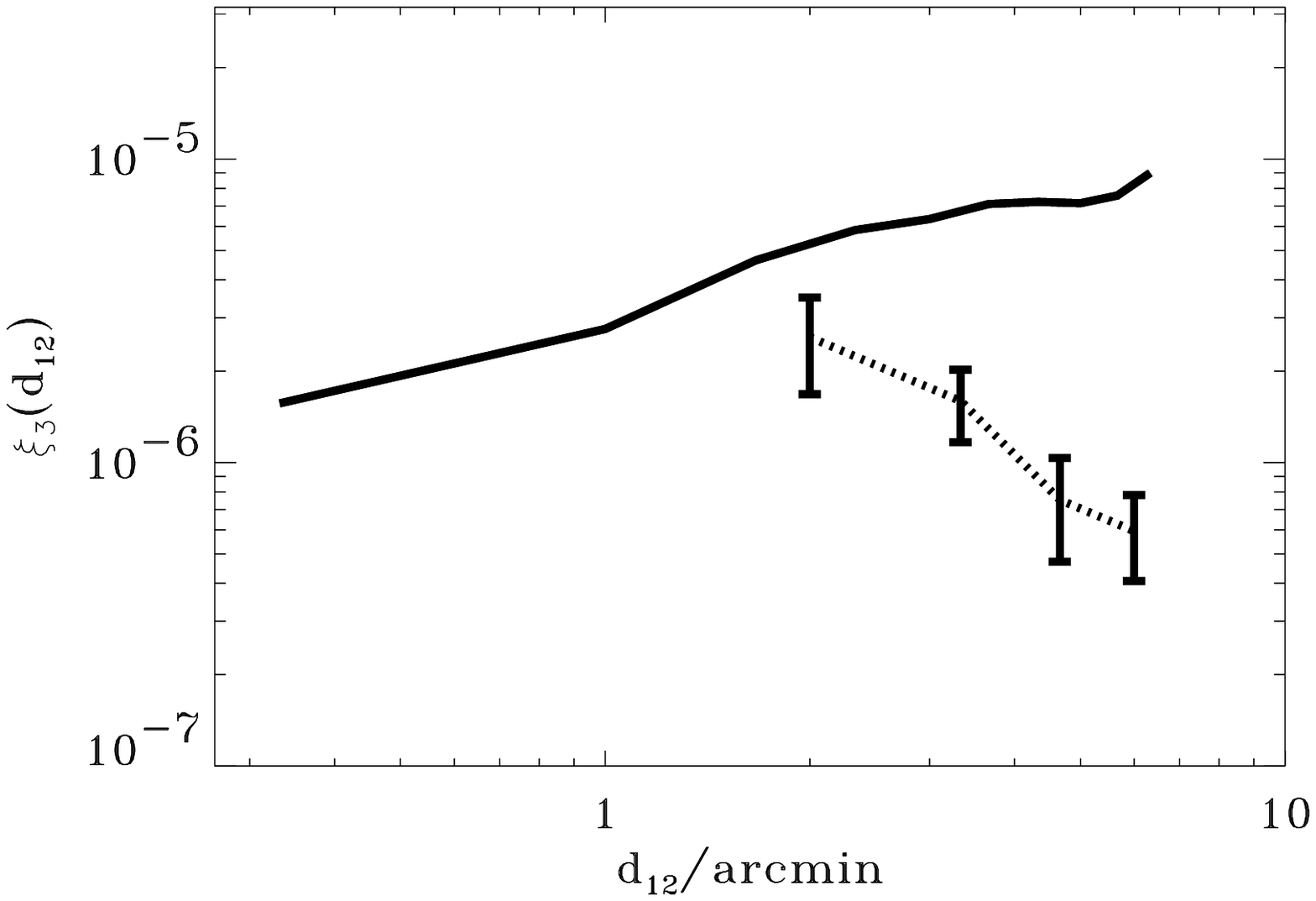,height=5.cm}
}
\caption[]{On the left, results for $\overline{\xi_3}(d_{12})/
\overline{\xi_2}(d_{12})^2$
for the VIRMOS-DESCART survey (dot-dashed
lines: $E-B$ mode for the 2-points function, solid line: $E+B$ mode
for the 2-points function). This is
compared to $\tau $CDM and OCDM results (dotted and dashed lines
respectively). Right plot: dashed line is $\overline{\xi_3}(d_{12})$
for the VIRMOS-DESCART survey, compared to the same quantity measured on
the stars.
\label{xi3.ps}}
\end{figure}
Figure \ref{xi3.ps} shows the result on the VIRMOS-DESCART survey. The treatment
of the $B$ mode is still uncertain, and the redshift uncertainty still too large, which
makes very difficult the interpretation in terms of cosmological parameters. However
Figure \ref{xi3.ps} shows that the order of magnitude, and the slope of the signal
are consistent with the expectations. For instance, the signal from the stars before
PSF correction is completely different in shape and amplitude.

\begin{figure}
\centerline{
\psfig{figure=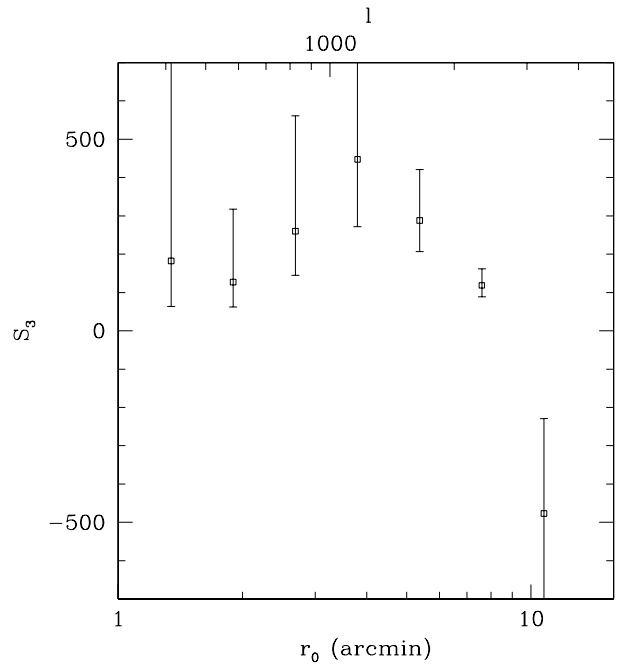,height=9.cm}
}
\caption[]{Skewness of the convergence as measured in Pen et al. (2002), on
the VIRMOS-DESCART survey. The overall significance of the measurement if
$3.3\;\sigma$, which was computed using Monte Carlo sampling of the errors
from ray-tracing simulations. A comparison of the signal with tese simulations
shows that $\Omega_0 < 0.4$ at the $90\%$ level.
\label{s3.eps}}
\end{figure}
In Pen et al. (2002), the idea is to compute the convergence aperture mass 3-points
function from an integral of the shear 3-points function. This solution avoids
the problem of drawing cells across a complex field geometry and 
   presents the advantage to
 estimate  the third moment of the convergence $\kappa$, which is the field
of physical interest. Unfortunately, its measurement is still very noisy, 
   because it uses
a compensated filter that  removes the low frequency modes for
any target frequency (which is not the case for a top-hat filtering). The resulting
skewness is shown on Figure \ref{s3.eps}, and is consistent with
$\Omega_0 < 0.4$ at the $90\%$ level.

Other approaches have been proposed (Zaldarriaga \& Scoccimarro 2002, Schneider
\& Lombardi 2003, 
Takada \& Jain 2003) which all deal with trying to optimize the signal-to-noise
by looking for the best galaxies triangle configurations
containing the highest signal. They have not yet been applied to the data.

\section{Galaxy biasing}

A direct byproduct of cosmic shear observations is the measure of the
light/mass relation, the so-called biasing parameter $b$ defined as the ratio of
the galaxy density contrast to the dark matter density contrast

\begin{equation}
\delta_{\rm gal}=b\;\delta_{\rm mass}.
\label{simpleb}
\end{equation}
This is in fact a highly simplified model, which assumes that the biasing
does not vary with scale and redshift, and that the relation between mass and light
is deterministic. While in the real world, none of these assumptions are 
   correct, 
this model has the advantage to be tractable analytically, and to 
    provide an average biasing
estimates, which is still very useful.
Nevertheless, it is possible to go beyond this simple model by combining a measurement
of the dark matter clustering, galaxy clustering, and their cross-correlation by defining
a biasing $b$ and cross-correlation $r$ such that:

\begin{equation}
b={\langle N_{\rm ap}^2\rangle \over \langle M_{\rm ap}^2\rangle} ; \; \;
r={\langle M_{\rm ap}N_{\rm ap}\rangle \over \langle N_{\rm ap}^2\rangle^{1/2}
\langle M_{\rm ap}^2\rangle^{1/2}},
\label{brdef}
\end{equation}
\begin{figure}
\centerline{
\psfig{figure=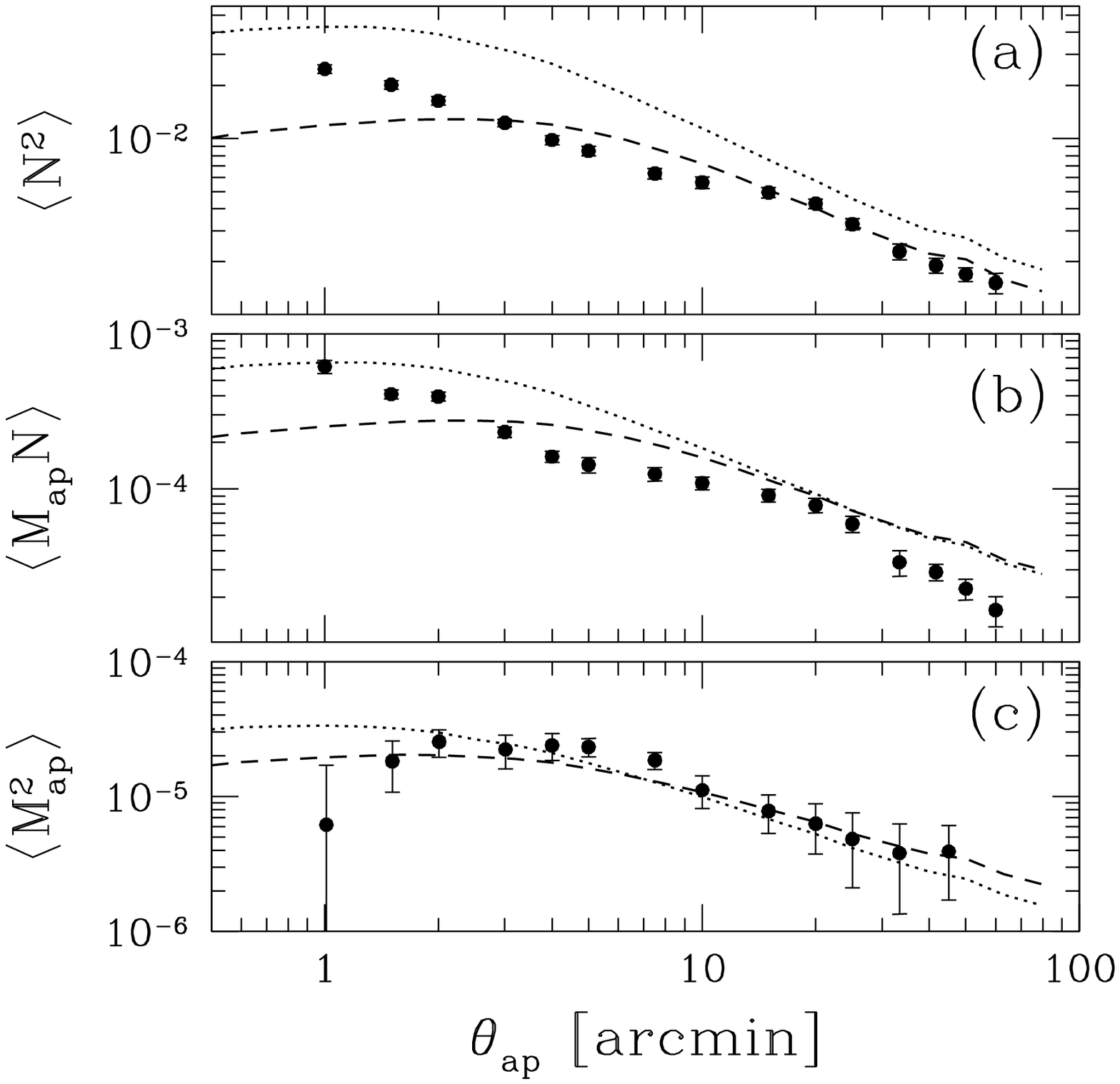,height=7.cm}
\psfig{figure=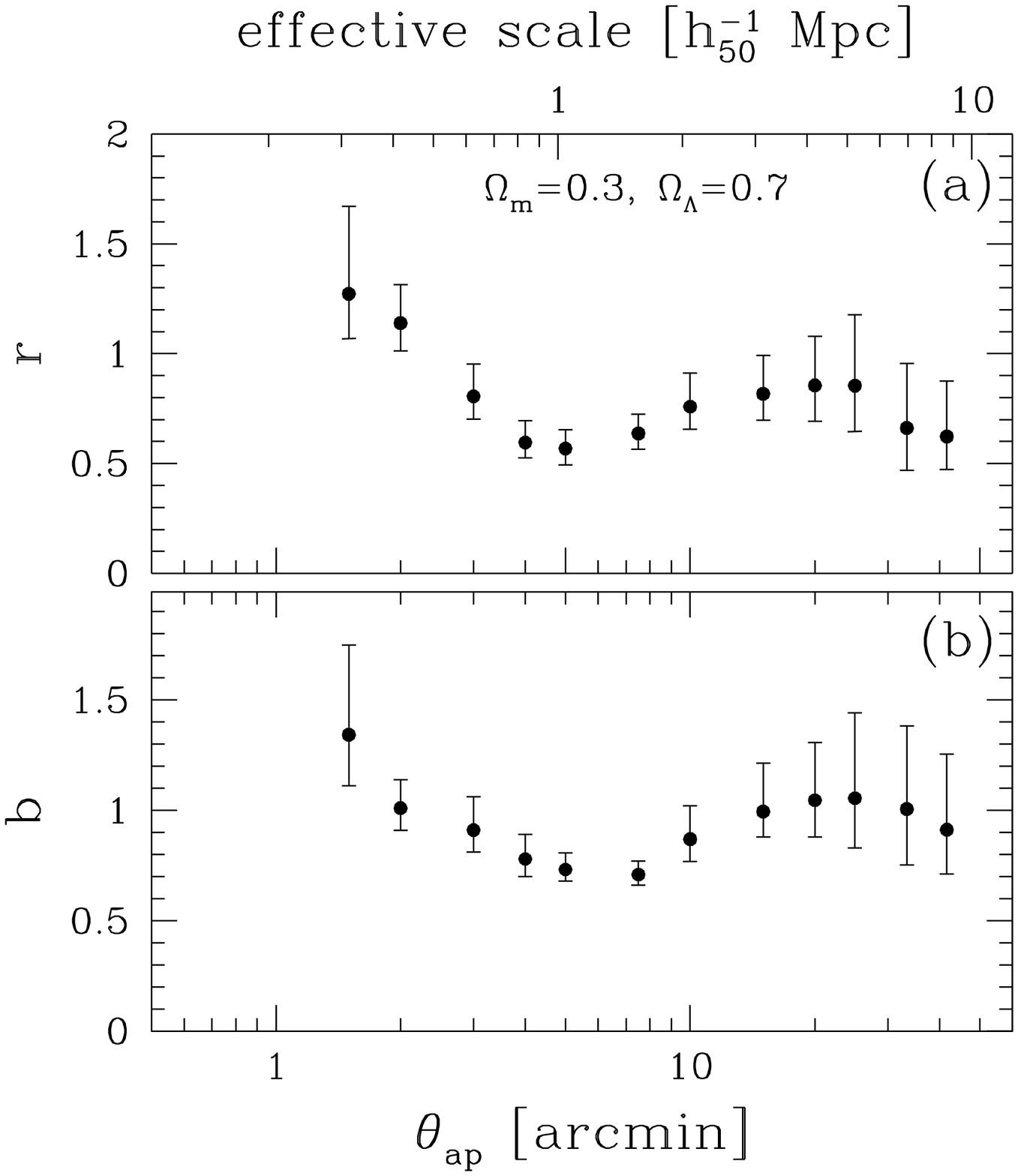,height=7.cm}
}
\caption[]{{\bf Left plot}: The measurements of $\langle {\cal
N}^2\rangle$ (panel~a), and $\langle {\cal N}M_{\rm ap}\rangle$
(panel~b) as a function of angular scale from the RCS data. Panel~c
shows $\langle M_{\rm ap}^2\rangle$ as a function of angular scale
from the VIRMOS-DESCART data. The
error bars for $\langle M_{\rm ap}^2\rangle$ have been increased
to account for the unknown correction for the observed ``B''-mode.
For reference, a few models have been plotted, 
assuming $b=1$ and $r=1$, for an OCDM cosmology
(dotted line; $\Omega_m=0.3$, $\Omega_\Lambda=0$, $\sigma_8=0.9$, and
$\Gamma=0.21$) and a $\Lambda$CDM cosmology (dashed line;
$\Omega_m=0.3$, $\Omega_\Lambda=0.7$, $\sigma_8=0.9$, and
$\Gamma=0.21$).  Note that the points at different scales are only
slightly correlated. {\bf Right plot}: The measured value of the galaxy-mass
cross correlation coefficient $r$ as a function of scale for the
$\Lambda$CDM cosmology. (b) The bias parameter $b$ as a function
of scale. The upper axis indicates the effective physical scale probed by the
compensated filter at the median redshift of the lenses
$(z=0.35)$.
\label{biasing.ps}}
\end{figure}
where $N_{\rm ap}$ is the galaxy number count density contrast smoothed 
   with a compensated
filter. Therefore, $N_{\rm ap}$ is similar to $M_{\rm ap}$, except that 
    it applies to
the number count instead to the shear. As we discussed before, the compensated 
    filter
is a passband filter, quite narrow in Fourier space. If one chooses
the number count fluctuations  $N_{\rm ap}$ to be a foreground 
   galaxy population with a 
narrow redshift distribution, then the biasing and cross-correlation $b$ 
    and $r$
emerging from Eq(\ref{brdef}) will be relatively localized in redshift AND wavelength.
The combinaison of well localized redshift and wavelength corresponds to a roughly
fixed physical distance.
Therefore we can say that, even with the simple scheme of galaxy biasing
given by Eq(\ref{simpleb}), an estimate of $b$ and $r$ from Eq(\ref{brdef})
is fairly local in physical scale, for the foreground galaxy population
under consideration (Schneider 1998, Van Waerbeke 1998). This result has been
proved to be robust against a wide range of cosmological parameters and power
spectra (Van Waerbeke 1998).

This idea has been applied for the first time in the RCS data (Hoekstra et al. 2001).
Unfortunately, this survey is not deep enough to provide an accurate measure of
the dark matter clustering that could allow to separate $b$ and $r$. Instead, the
authors measured the ratio $b/r=1.05^{+0.12}_{-0.10}$ for the favored
$\Lambda$CDM model ($\Omega_0=0.3$ and $\Omega_\Lambda=0.7$). On the other hand,
a combination of deep and shallow survey could help to measure the bias and
the cross-correlation independently. This was done by combining the RCS and
VIRMOS-DESCART surveys (Hoekstra et al., 2002). RCS is a wide shallow survey with a mean
source redshift of $\sim 0.6$, and VIRMOS-DESCART is a deep survey with a mean source redshift
$\sim 0.9$. By selecting the foreground population with a median redshift
$\sim 0.35$ on the RCS survey, the number counts
$\langle N_{\rm ap}^2\rangle$, and the cross-correlation
$\langle M_{\rm ap}N_{\rm ap}\rangle$ were measured. The aperture mass
$\langle M_{\rm ap}^2\rangle$ is measured on the deep survey. Figure \ref{biasing.ps}
shows the measured $b$ and $r$ as a function of scale (angular scales are also converted
to physical scale for a given cosmological model, with the lenses at $z=0.35$). Although
a proper interpretation of the measurement requires a better knowledge of the redshift
distribution and cosmological parameters, it is a direct indication of the
stochasticity ($r<1$) of the biasing at small scale, and that the biasing varies with
scale as we approach the galactic scales, below $1'$. The foreground galaxies
were selected in $R$, and it was found that $b=0.71^{+0.06}_{-0.04}$ on a scale
$1-2\;h^{-1}_{50}\;{\rm Mpc}$, and $r$ reaches a miminum value of $r=0.57^{+0.08}_{-0.07}$,
at $1\;h^{-1}_{50}\;{\rm Mpc}$. We should note that $b$ tends toward $1$ at larger
scale.

\section{Dark matter power spectrum inversion}

The central interest in cosmic shear observation is dark matter. This is probably
even more important than measuring the cosmological parameters, for which we have
some hope to measure them very accurately in the future (although there is the issue
of degeneracies where lensing can help). One important question is then: what
can we say about the dark matter distribution, provided we know all the cosmological
parameters? This is nothing else but to try to map the dark matter in the same
way we map the galaxies or the cosmic microwave background, or at least to
measure its power spectrum in three dimensions, for all possible scales, independently
of any evolution model. This is in principle possible from a direct
inversion of Eq(\ref{pofkappa}), but there are two issues here. One is that
virtually, all physical wavelengths $k$ are projected out to
give a single angular wavelength $s$, and with a naive
deprojection, one needs some cut-off somewhere in $k$-space
to perform the invertion.
The other issue is that the 3D power spectrum evolves non-linearly with redshift
in the non-linear scales, therefore how could we be independent of any modeling when
inverting the 2D convergence power? The first 2D convergence power spectrum
estimate was performed in Pen et al. (2002) on the VIRMOS-DESCART data, and
in Brown et al. (2003) on the COMBO-17 data, but the spectrum inversion
was not done.

\begin{figure}
\centerline{\vbox{
\psfig{figure=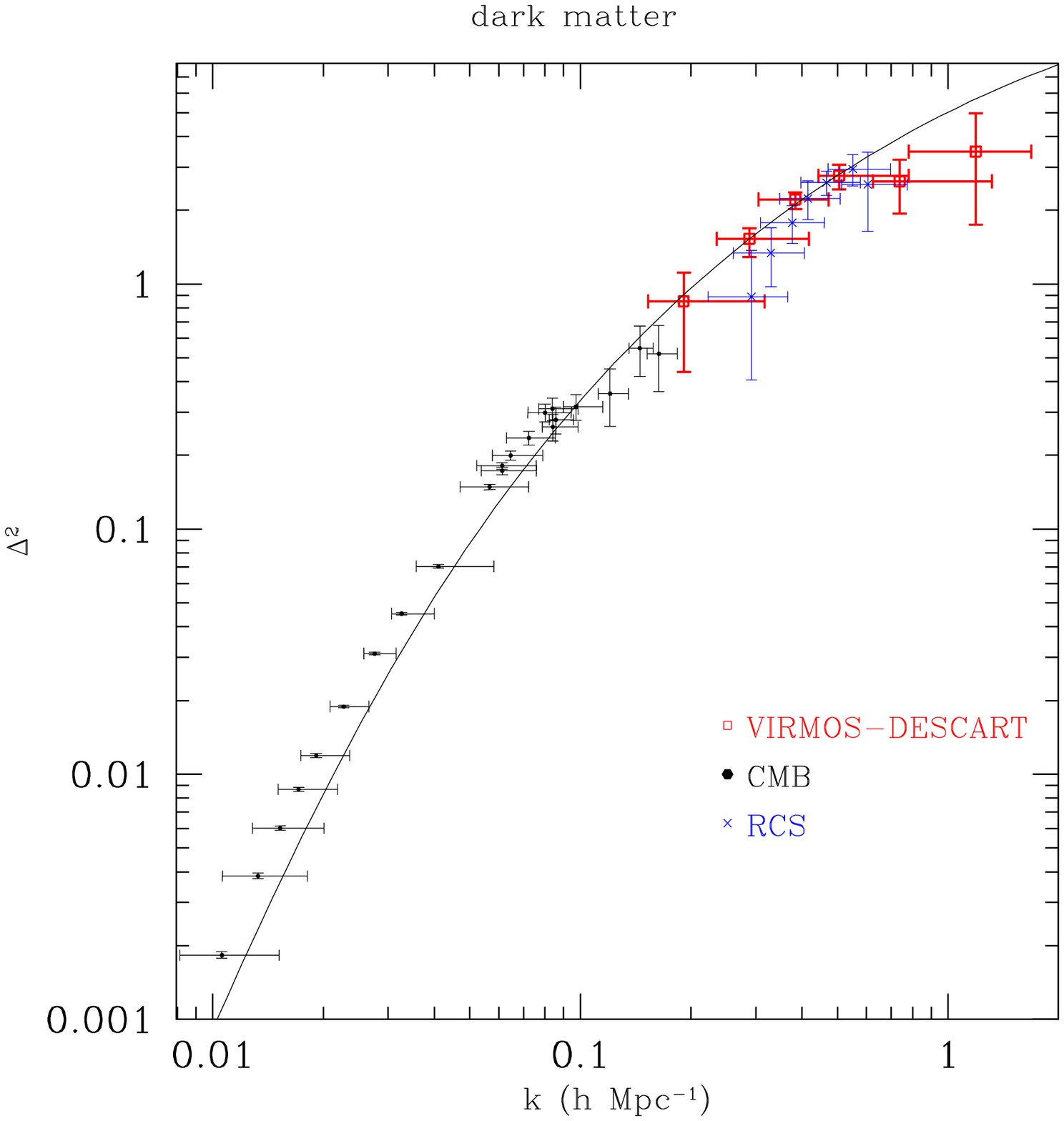,height=10.cm}
\psfig{figure=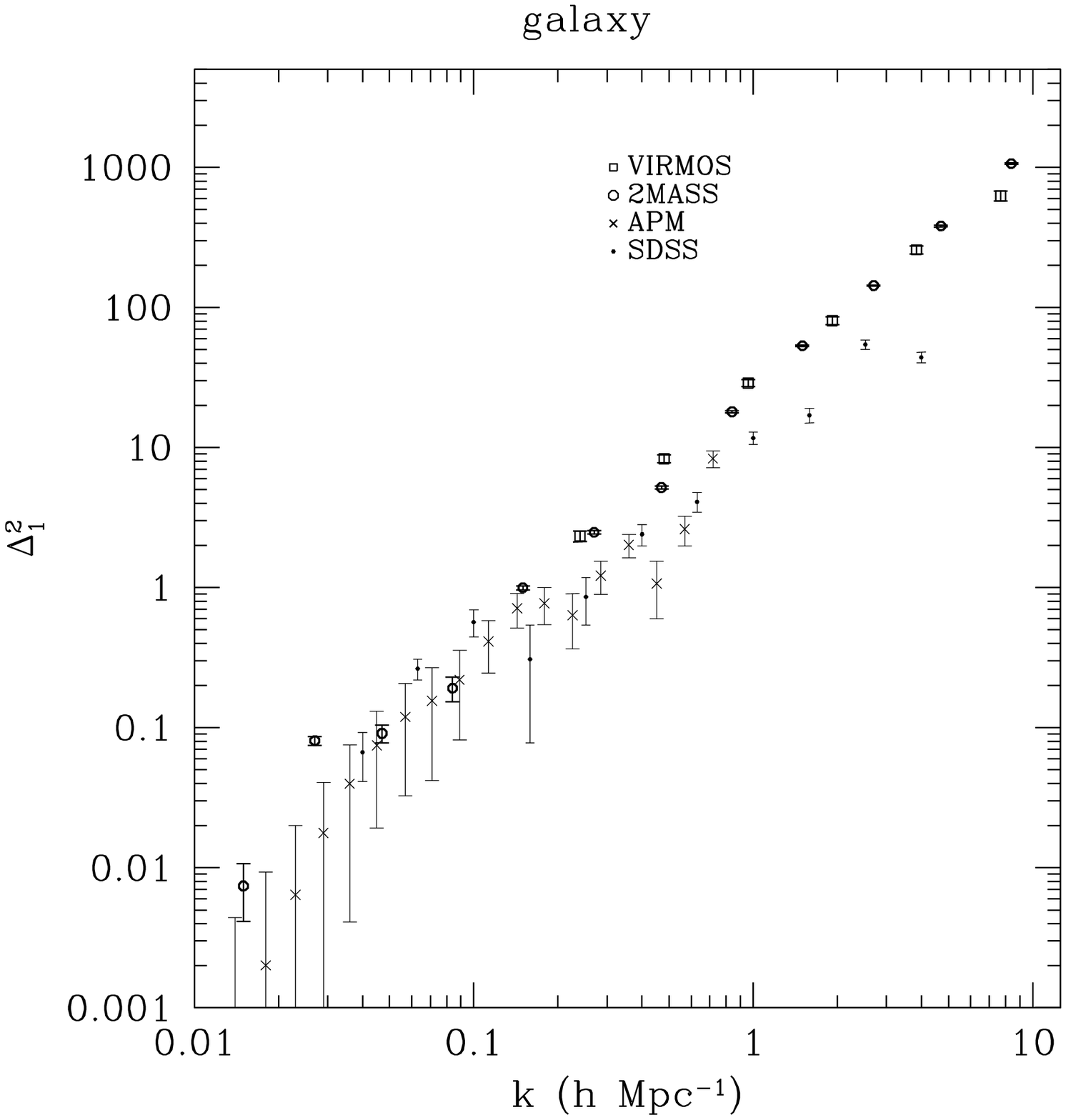,height=10.cm}
}}
\caption[]{{\bf Top}: Dark matter 3D power spectrum, deprojected from the
2D convergence power spectrum measured in the VIRMOS survey,
using SVD (Pen et al. 2003). The power is
rescaled to $z=0$, points are compared to the CMB (WMAP) and RCS lens survey.
{\bf Bottom}:
Galaxy 3D power spectrum deprojected using the same method. For comparison,
points from 2MASS, APM and SDSS are also shown.
\label{deproj.ps}}
\end{figure}
Pen et al. (2003) investigated the inversion using a singular decomposition
technique, an extension of the minimum variance estimator deprojection
developped in Seljak (1998). The non-linear evolution of the 3-D power
spectrum was assumed to evolve {\it linearly} with redshift even in the non-linear
regime. This hypothesis is, surprisingly, a viable assumption within the scale range
of interest, and produces errors still smaller than the statistical errors.
The result is shown on Figure \ref{deproj.ps} for the dark matter (top)
and the galaxies (bottom). It shows a very nice agreement with the cosmic microwave
background $C_l$'s (WMAP points extrapolated at $z=0$, see Spergel et al. 2003),
and with clustering measurements
from other galaxy surveys.

A dark matter-galaxy cross-correlation was also deprojected, allowing
Pen et al. (2003) to estimate the 3D biasing $b$ and matter-light correlation $r$.
They found $b=1.33\pm0.19$ and $r=0.68\pm0.24$ for the $I$-selected galaxies.
The bias value is slightly different than the one measured from the aperture mass
on the RCS survey
(section 5), but we should keep in mind that the galaxy populations are
different ($R$ compared to $I$ selected galaxies for the RCS and VIRMOS surveys
respectively). The physical scales probed in VIRMOS are also larger
because it is a deeper survey than in RCS.

\begin{figure}
\centerline{
\psfig{figure=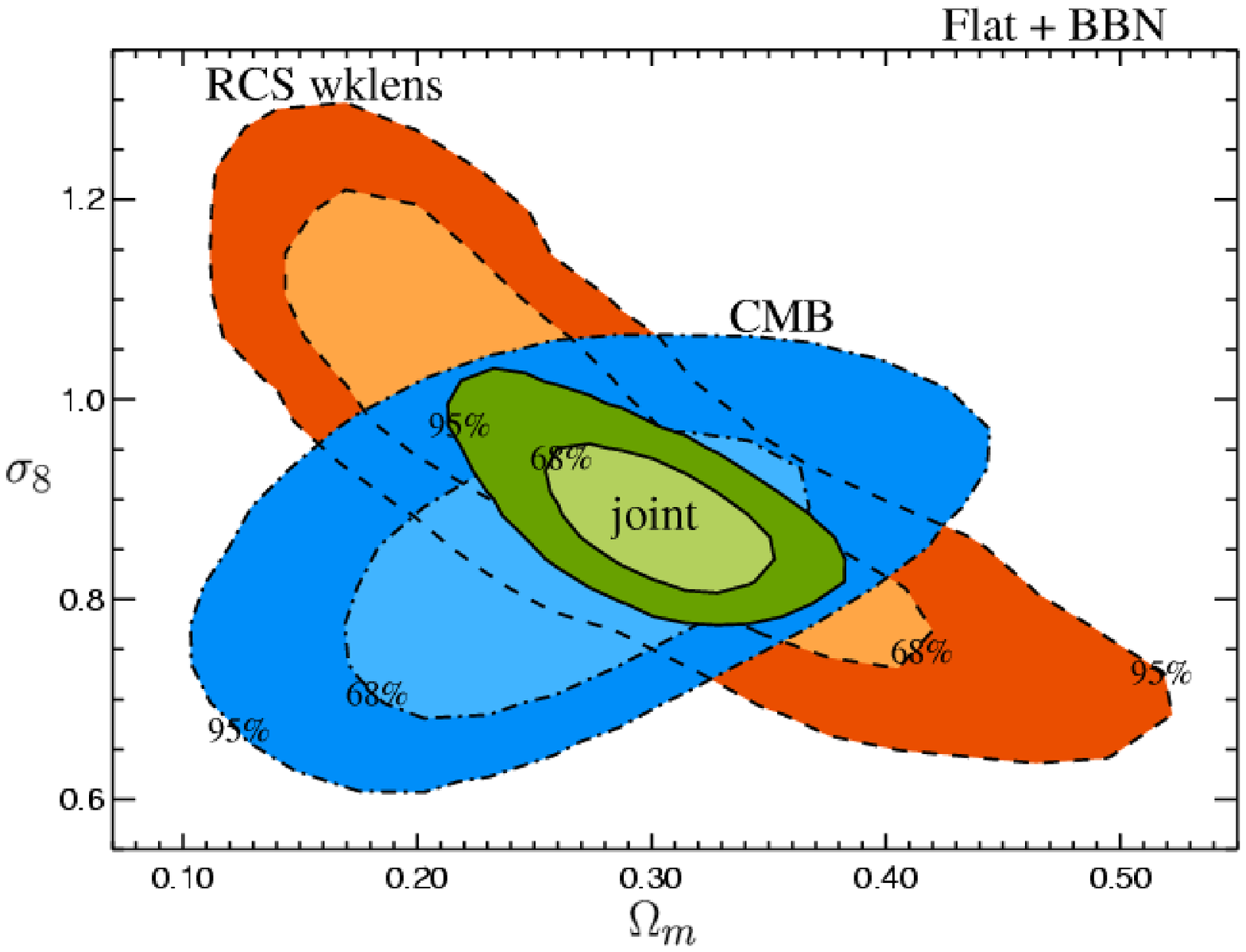,height=7.cm}
}
\caption[]{The two dimensional, marginalized
likelihoods for the $(\Omega_m,\sigma_8)$ plane. The overlaid, filled
contours show the 68\% and 95\% integration levels for the
distributions. Bottom -- RCS only, Middle -- CMB only, Top --
CMB+RCS. Courtesy Contaldi et al. 2003.
\label{lenscmb.ps}}
\end{figure}

\section{Gravitational Lensing and Cosmic Microwave Background}

The use of lensing with other experiments improves the
accuracy of cosmological parameter measurements and eventually breaks some
 intrinsic degeneracies attached to each. The potential interest 
 of combining lensing by large scale structures and cosmic microwave background
experiments has been studied in Hu \& Tegmark (1999). The joint study of the
weak lensing RCS survey and the WMAP data  performed in
Contaldi et al. (2003) is shown on Figure \ref{lenscmb.ps}
  and illustrates the gain
of this combination: it provides a direct evidence of the low value of
the matter density $\Omega_0$, which indicates a high non-zero value
for the cosmological constant, independently of the supernovae
result.

\section{Approximations and Limitations}

\subsubsection{Born approximation and lens-lens coupling}
The lensing theory developed in Section 1 assumes  the lens can be projected
onto a single plane, and therefore that the ray-tracing through a thick
lens is equivalent to a thin lens appropriately weighted. 
  As it  has been quantified by Bernardeau et al. (1997),
Schneider et al. (1998) or Van Waerbeke et al. (2002), it turns out to be 
 a very good approximation. If we call $\thetag$ the
direction of the unperturbed ray trajectory, 
 a ray-light passing through a first lens will
be slightly deflected by an angle $\deltag\thetag$, and will impact the
second lens at a position angle $\thetag+\deltag\thetag$ instead of
$\thetag$ if the light ray were unperturbed. From a 
  perturbative point of view, it means that
 expression Eq(\ref{amplidef}) has a correction term because   
the position angle to compute the lens strength is no
longer $\xg=f_K(w)\thetag$, but

\begin{equation}
x_i=f_K(w)\theta_i-{2\over c^2}\int_0^w
{\rm d} w'\;f_K(w-w')\, \partial_i \Phi^{(1)}(f_K(w)\thetag,w').
\end{equation}
Eq(\ref{amplidef}) is therefore replaced by
${\cal A}_{ij}(\thetag)=\delta_{ij}+{\cal A}^{(1)}_{ij}(\thetag)+{\cal A}^{(2)}_{ij}(\thetag)$
with

\begin{eqnarray}
{\cal A}_{ij}^{(2)}(\thetag,w)&=&-{2\over c^2}\int_0^w {\rm d} w'\;{f_K(w-w')
f_K(w')\over f_K(w)} \cr
&\times & \left[\Phi_{,ikl}(f_K(w')\thetag,w')\,
x_l^{(1)}(\thetag,w')\delta_{kj}
+\Phi_{,ik}(f_K(w')\vc\theta,w'){\cal A}_{kl}^{(1)}(\thetag,w') \right]. \cr
\label{born}
\end{eqnarray}
Given that the correction to the light trajectory is a second order
effect in the perturbation, it is expected to become important in any high
order statistics of the lensing fields. Mathematically, indeed, they have the
same order than the second order dynamical correction (which is proportional to
the second order gravitational potential $\Phi^{(2)}$). It turns out that the
light trajectory correction is much smaller than the dynamical second
order correction. The reason is that Eq(\ref{born}) involves a second
lensing efficiency factor (the ratio of angular diameter distances $f_K$'s),
which is not present in the second order dynamical correction.

\begin{figure}
\centerline{
\psfig{figure=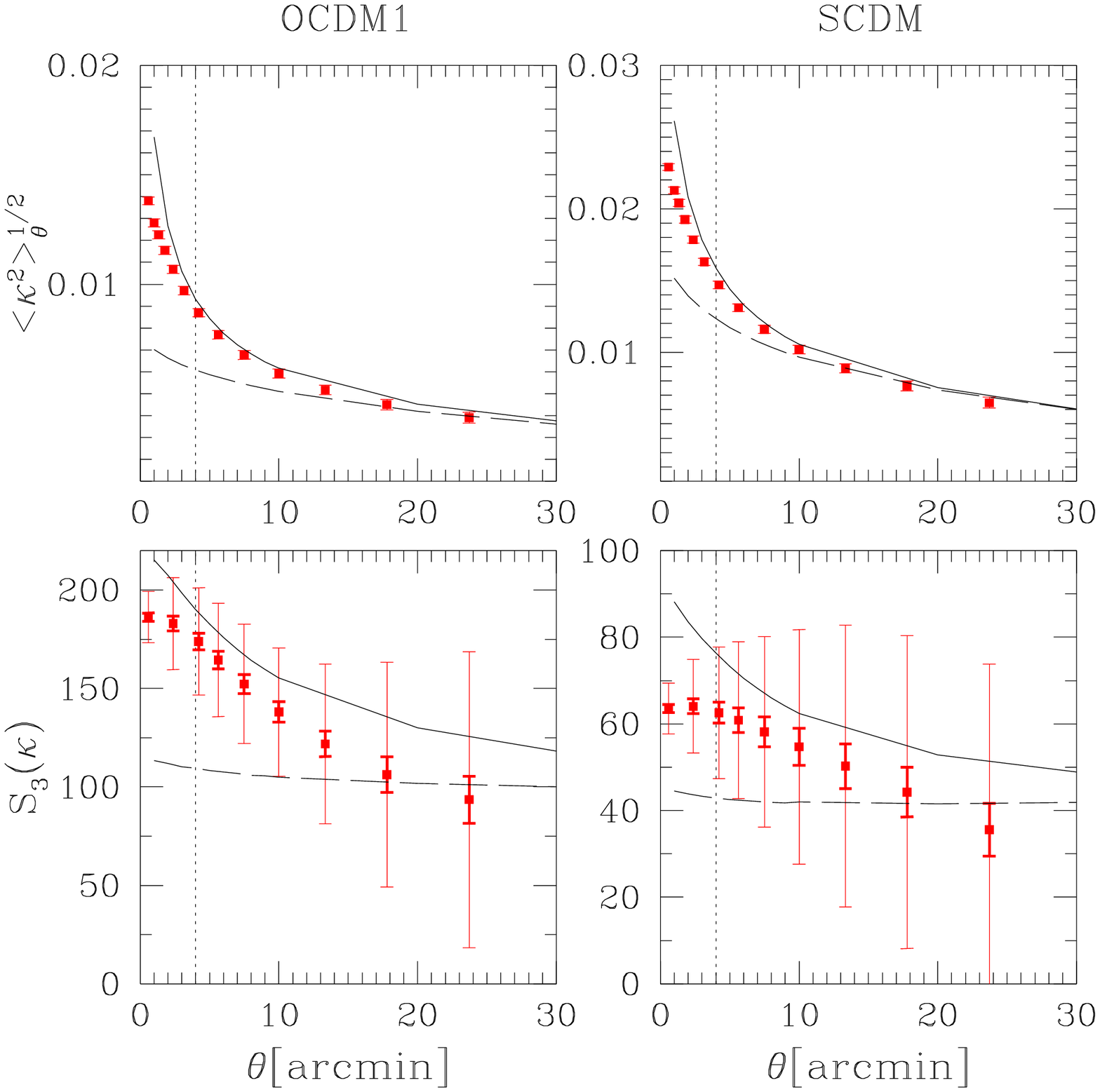,height=6.9cm}
\psfig{figure=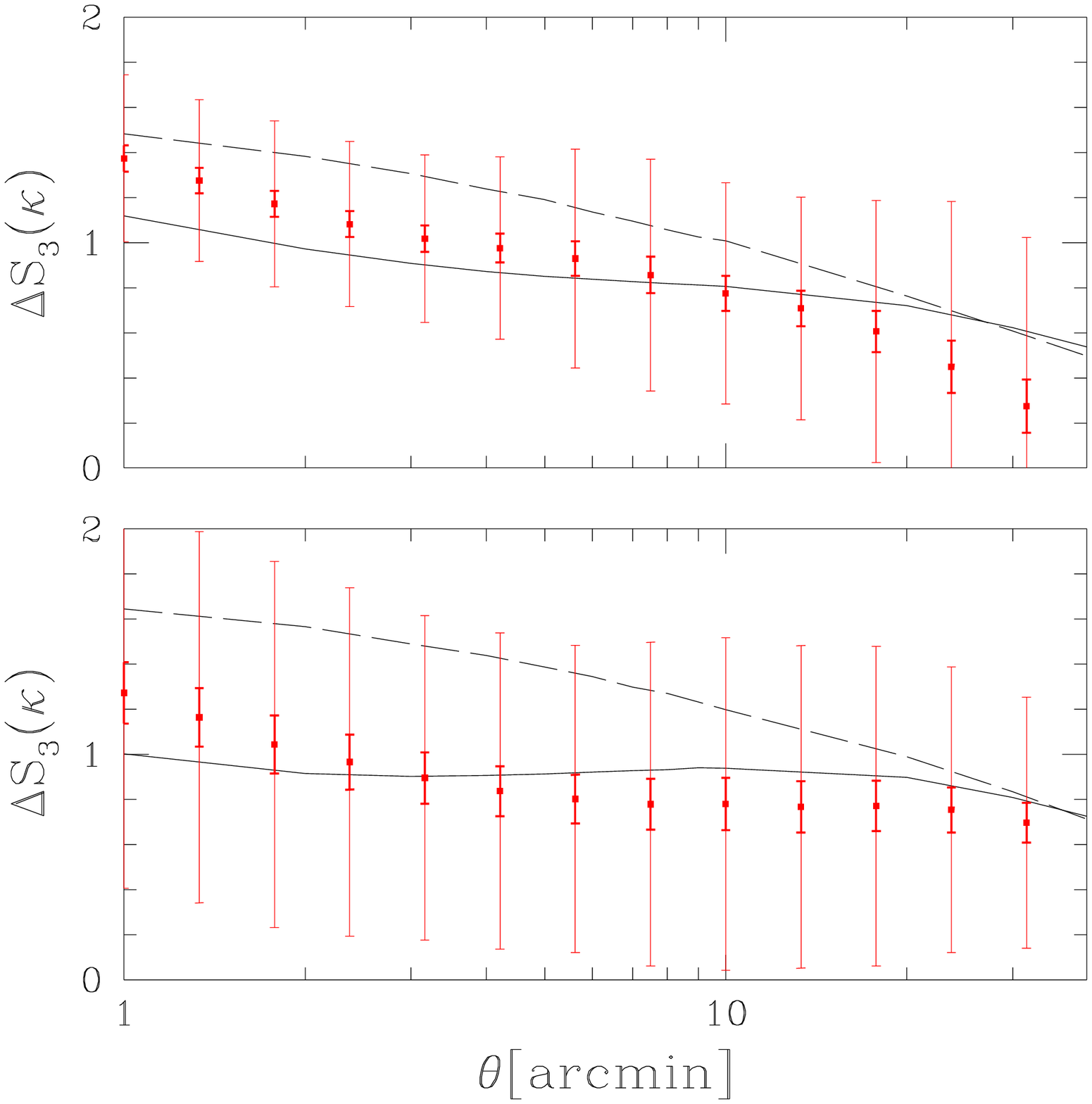,height=5.cm}
}
\caption[]{{\bf Left panel}: variance and skewness
of the top-hat smoothed convergence field
a OCDM ($\Omega_0=0.3$, $\sigma_8=0.85$, $z_s=1.03$, $\Gamma=0.21$)
and SCDM model ($\Omega_0=1.$, $\sigma_8=0.6$, $z_s=1.03$,
$\Gamma=0.5$). The solid lines show the
non-linear predictions, and the dashed lines the leading order of
the perturbation theory calculations. The
vertical dotted lines in the left panels denote the reliable scale
limit fixed by the resolution of the ray-tracing simulation.
The ``large'' error bars correspond to a survey of $25$ square degrees,
and the small error bars to $1000$ square degrees.
{\bf Right panel}: Skewness correction due to the Born approximation
and lens-lens coupling
for SCDM (top) and OCDM1 (bottom)
measured in ray-tracing simulation.
\label{bornapprox.ps}}
\end{figure}
Figure \ref{bornapprox.ps} shows several comparisons of the non-linear
prediction for the second and third order statistics with a measurement
of the same statistics done in ray-tracing simulations. It clearly
demonstrates that non-linear calculations give quite accurate results, and
that approximations related to the ray trajectories are valid to better than
$2\%$.

\subsubsection{Non-linear lensing effects}
To first approximation, we consider the galaxy ellipticity an unbiased estimate
of the shear. However, Eq(\ref{distodef}) tells us that lensing is
really non-linear. This approximation has been estimated in
Barber (2002): it is negligible for sources at redshift less than $z \simeq 1$ and
for scales larger than $5'$, while at smaller scale, a few percents effects 
  could be
 detected.  Fortunately, the use of the full non-linear lensing equation
 does not present any theoretical or technical difficulties, so 
   small scale non linear lensing effect can be easily handled. 
   It is just usually ignored in most of the theoretical
and numerical works.

\subsubsection{Non-linear modeling}
On the other hand, Figure \ref{bornapprox.ps} also demonstrates that 
  the accuracy of non-linear
predictions on the 2-points statistics is never better  
better than $10\%$, while it is never better than $\sim20-30\%$ for the 
   skewness. This
theoretical limit is a severe issue (Van Waerbeke et al. 2002) since 
  one cannot
 expect to do precision cosmology if the accuracy of the 
 model we use to extract the cosmological
parameters is worse than the precision we want to reach on the
cosmological parameters (that is a few percents). Smith et al. (2003) proposed
an improved version of non-linear modeling, which is unfortunately still
insufficient. In particular, to increase the precision, we still do not know
whether the baryons have to be taken into account in the modeling or not. The goal
here is a modeling accurate to $1-3\%$, if one wants to reach the same accuracy
on the cosmological parameters.

\subsubsection{Intrinsic alignment}
Gravitational lensing is not the only natural process which produces 
   alignment of
galaxies over large distances. Intrinsic alignment might occur from tidal
fields, and produce galaxy shape correlations over cosmological distances, and
contaminate cosmological signal (Croft \& Metzler 2001, Catelan \& Porciani 2001,
Heavens et al. 2000, Catelan et al. 2001, Hatton \& Ninin 2001)
which should
in principle split, in a predictable way, into $E$ and $B$ modes. There is
unfortunately only partial agreement between the different predictions.
Moreover, most of the predictions stand for dark matter halos, while we are
in fact observing galaxies, which should experience some alignment mixing.
This has not been simulated so far. Concerning the dark matter halos alignment,
despite the disagreement among the predictions, 
it is generally not believed to be higher than
a $10\%$ contamination for a lensing survey with a mean source
redshift at $z_s=1$. An exception is Jing (2002), who suggested
that intrinsic alignment
could dominate the cosmic shear even in deep surveys. This possibility
is already ruled out by observations:
this would indeed imply a very low $\sigma_8 \sim 0.1$ if the observed signal were
dominated by intrinsic alignment, and we should also observe
an increase of the effect as we go from deep to shallow survey, which is not
the case (see Figure \ref{mapplot}). In any case, intrinsic alignment
contamination {\it might} be an issue for studies using a single source redshift in
their analysis. In the future, this will not be the case since photometric redshifts
will be available. In that case, the effect can be suppressed
by measuring the cosmic shear correlation
between distant redshift sources, instead of measuring the fully projected signal.
Consequently, intrinsic alignment should not be considered as a critical
issue (Heymans \& Heavens 2003, King \& Schneider 2003).
Pen et al. (2000) and Brown et al. (2002) reported the first two evidences for
intrinsic alignment in the nearby Universe, which are not too inconsistent with
the predictions.

\subsubsection{Source clustering}

Source clustering arises because a subset of sources overlap with a subset of the
lenses which are probed. There is therefore a natural bias to measure the signal
preferentially
in high density regions, across the overlap area. This effect gives rise to
correction terms in high order statistics (Bernardeau 1998). It is easy to
understand the problem if we model the source redshift distribution including
a clustering term:

\begin{equation}
p_w(\thetag,w)={\bar p}_w(w)(1+\delta_{gal}(f_K(w)\thetag,w)+...),
\end{equation}
which replace the source redshift distribution $p_w(w)$ in Eq(\ref{gfunc}).
It is then easy to see that a density coupling occurs in Eq(\ref{kappadef}).
The source clustering effect was extensively studied by Hamana et al. (2002).
They confirmed that it is not an issue for the 2-points statistics, but
could be as high as $10-20\%$ for the skewness of the convergence,
for a narrow redshift distribution. In case of the broad redshift distribution,
the effect is diluted by the bulk of non-overlapping areas. For future surveys,
an accurate measure of the high order statistics will require a precise
estimation of this effect, which is not a problem by itself, but it
must be done. Schneider et al. (2003) predicted that source clustering could
produce $B$ mode in the shear maps, which has not being tested against
reay-tracing simulation yet. However, the predicted amplitude is 2 orders of
magnitude and below for aperture mass scales larger than $1'$.

\subsubsection{PSF correction}
With the non-linear modeling of the power at small scale, this is certainly
the most serious issue concerning the cosmological interpretation of the cosmic
shear signal. Again, if we want to reach a few percents accuracy on
cosmological parameters measurements, we need a PSF correction with that
accuracy. So far we are able to reach $10\%$ precision with the KSB method
for a typical signal measured on sources at $z=1$ (Erben et al. 2001,
Bacon et al. 2001,
Van Waerbeke et al. 2002). This is reasonably good, but we still need to gain
one order of magnitude (in addition to the order of magnitude we need to
gain for the non-linear modeling as well). The $10\%$ uncertainty is an upper
limit, which comes
from the large $B$ mode found in all surveys, at different scales,
for probably different reasons (for instance, RCS have $B$ mode at small scale
only they may have measured intrinsic alignment). This upper limit is reduced if one
uses the scales with very small or no $B$ mode, but then some cosmological
information is lost. So far, our understanding
of the PSF modeling is insufficient in particular concerning the PSF variation
(and stability) accros the CCD's and the contribution of high frequency modes.
Space data are often viewed as potentially
systematics-free. This is unfortunately not true, since all space data which have been
processed for cosmic shear, required a significant PSF correction. However,
the main difference between space and ground
based data is that, in space, the PSF is certainly more stable between exposures.
But one should not forget that in space, the PSF is $100\%$ instrumental
(it is the Airy spot, which is larger than the Airy spot on the ground because
space telescopes are smaller), and not atmospheric at all (which it is with ground
based data with larger telescopes). Dealing with a non circular
Airy spot to correct for the galaxy shapes was not trivial for the
Hubble Space Telescope for
instance, mainly because of the severe undersampling of the PSF (Hoekstra
et al. 1998). There is no intensive simulation of shear measurement under
various realistic
space image conditions, only qualitative estimations have been
done (R\'efr\'egier et al. 2003, Massey et al. 2003), which seems promising.

Finally, one should emphasize that the most difficult part of the PSF correction
is not the anisotropic correction, which is done quite accurately,
but the isotropic correction (Erben et al. 2001, Hirata \& Seljak 2003).
The ultimate limit of PSF correction in space and on the ground
is still an open question.

\section{Prospects}

\begin{figure}
\centerline{\vbox{
\psfig{figure=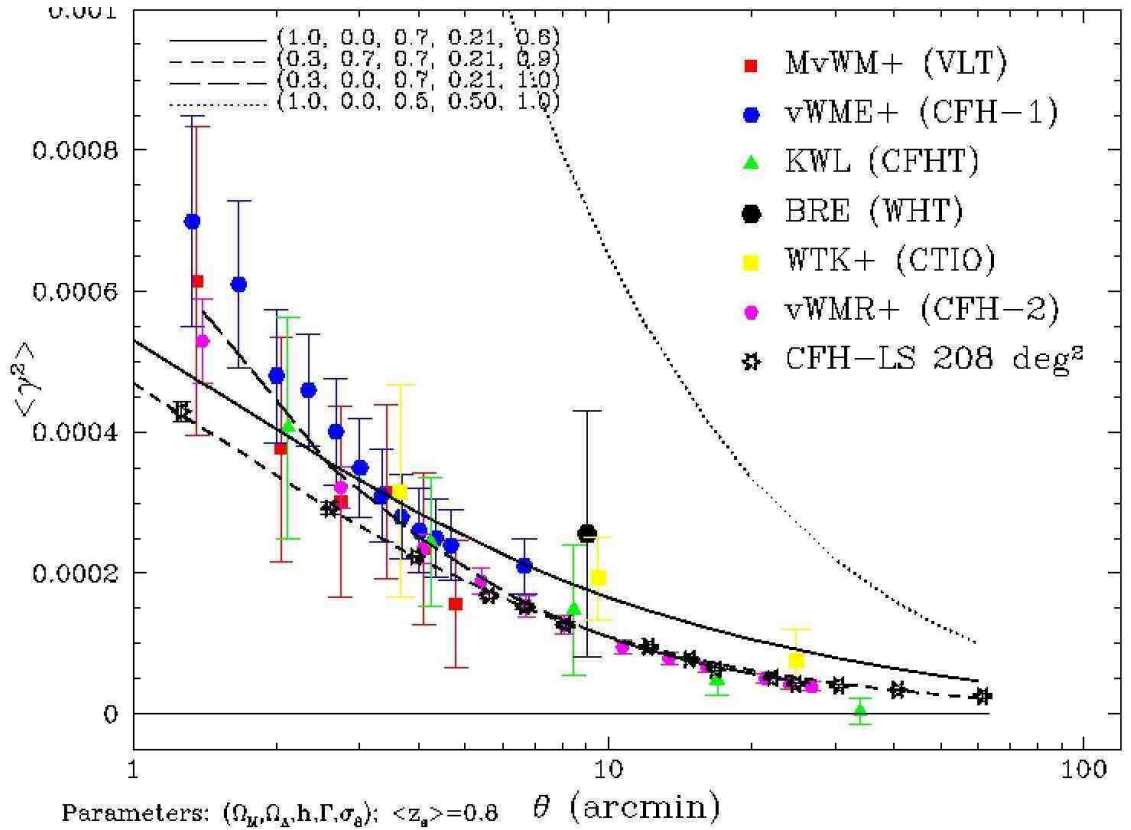,height=11.cm}
}}
\caption[]{Top hat variance of shear as function of angular scale from
  6 cosmic shear surveys. The  open
black stars are the predictions for the 
  CFHT-LS
  which will start by 2003 with Megacam at CFHT. This is
 the expected signal from  the ``Wide Survey'' which
  will cover 170 deg$^2$ up to $I_{AB}=24.5$. For
most points the errors are smaller than the stars.
\label{prospect1.ps}}
\end{figure}
\begin{figure}
\centerline{\vbox{
\psfig{figure=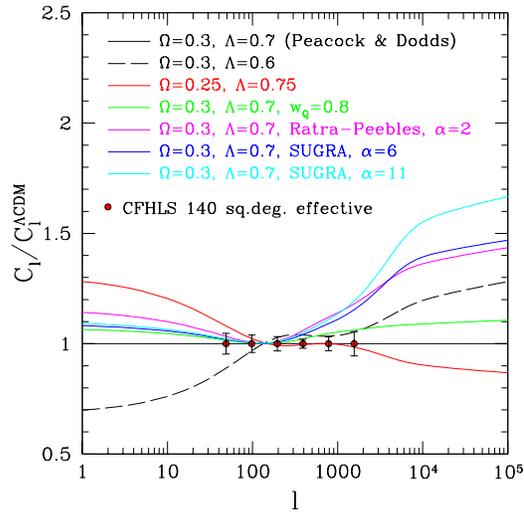,height=7.cm}
}}
\caption[]{Theoretical expectations on cosmological models 
 beyond the standard model from 
the wide CFHT Legacy Survey that will cover 170 deg$^2$. The dots with
error
bars are the expected measurements of cosmic shear CFHTLS data. The lines
shows various models
  discussed by Benabed \& Bernardeau (2001).
\label{prospect2.ps}}
\end{figure}

In the WMAP (Spergel et al. 2003) context, one can wonder whether
future cosmic shear surveys can 
 still provide useful cosmological informations that would not
be available otherwise from CMB and SNIa experiments. The answer is 
  clearly yes because cosmic shear is the only way to directly probe
dark matter on scales that cannot  be directly probed by other techniques.  
  It can explore the properties of the dark matter
  together with luminous matter on quasi-linear and non-linear scales 
 where the complexity of physical processes make theoretical and
 numerical predictions among  
   the most challenging tasks for the next decade.  Further, because cosmic 
  shear is sensitive to the 
growth rate of perturbations integrated along the line-of-sight,
the additional redshift information provides a tool to
study the structure formation mechanism and
the clustering history with look-back time. It is the purpose of {\it
tomography} to study the 3D matter distribution by combining the lensing
effect with the redshift information of the sources (Hu 1999, Heavens 2003).
This clearly belongs to the prospective part, and was not discussed in the
review.
 
The future key scientific goals are therefore the reconstruction of the 
 3-dimension dark matter power spectrum as function of redshift, the 
 analysis of the properties of the relation between light and mass,
 and the study of the dark energy equation of state.
  As shown in this review, the scientific studies of the
  cosmic shear surveys done so far already show that there are 
neither conceptual nor technical barriers that hamper these goals 
  to be achieved quickly.  Tegmark \& Zaldariaga (2002)
 with 
  the RCS cosmic shear survey
 and Pen et al. (2003) with the VIRMOS-DESCART 
  cosmic shear surveys have demonstrated that the 3-D power spectrum of
 the dark matter can be reconstructed. Remarkably, their results extend 
 monotonically toward small
scales the dark matter power spectrum derived from CMB experiments.
Likewise, Hoekstra et al. (2001)
  Hoekstra et al. (2002),  and Pen et al. (2003), have shown that 
  the 
   properties of the biasing and the dark matter-galaxy cross
correlation  can already be analyzed with present-day surveys. Finally,
 Bernardeau et al. (2003) as well as Pen et al.
(2003) have shown that high order statistics are already 
  measurable
  from ground based data covering 10 deg$^2$, thus providing independent 
informations on cosmological models, with eventually some 
  important degeneracies broken.

Although the B-mode contamination is still an important technical 
  issue that may slow down the cosmic shear developments, in principle the 
  next large, deep and multi-color surveys
will be in position to 
    address questions 
  relevant for cosmology and fundamental physics with a high degree of
precision.
Such surveys, covering hundreds of degrees, with multi-bands data 
are about to start.  Among those, the wide CFHT Legacy
Survey\footnote{http://wow.cfht.Hawaii.edu/Science/CFHLS/} will cover
170 deg$^2$, spread over three uncorrelated fields, 
in 5 optical bands, and a fraction will be followed up 
later in J and K bands with the wide field infrared camera WIRCAM at
CFHT. Figures \ref{prospect1.ps} and \ref{prospect2.ps} show some 
predictions of CFHTLS. On Figure \ref{prospect1.ps} we simulated
the expected signal
to noise of the shear variance as function of angular scale for
a $\Lambda$CDM cosmology.  The error bars are much smaller than 
the VIRMOS-DESCART survey which has the same depth as CFHTLS.
On Figure \ref{prospect2.ps}, we compare the expectations
of the CFHT Legacy Survey angular power spectrum 
with the predictions of several theoretical quintessence fields models. It
shows that  200 deg$^2$ deep survey with multi-color informations 
to get redshift of sources, one can already interpret cosmological data  
beyond standard interpretations.  The CFHTLS will be of considerable
interest because one of the fields is also a target for the VMOS/VDDS
spectroscopic survey (Le F\`evre et al 2003), the XMM-LSS survey 
(Pierre et al 2001) and also the COSMOS
Treasury Survey that will be done by the HST/ACS instrument.  Hence, 
in addition to a complete description  of the redshift distribution of 
the CFHTLS galaxies, as well as of the X-ray clusters and active galaxies, 
high accuracy shape measurement of galaxies will be feasible. This 
HST/ACS data set attached to a subsample of CFHTLS data 
will permit to check the reliability of ground based PSF
corrected shear catalogs but also to extend the shear analysis on very 
small scales, 
down to the galactic dark halos scales.  Join together with CMB, 
we then expect to get by 2005 a complete view of the dark matter
power spectrum and the biasing
from Gigaparsecs down to kiloparsecs scales, as well as a detailed 
description of individual dark halo properties
and of the redshift distribution of lenses and sources
(Cooray \& Sheth, 2002). One should mention the extensive search for dark matter
halos in the Suprime-Cam fields (Miyazaki et al. 2003): this is
more a cluster physics-oriented, but it illustrates very well the future
developments with cosmic shear data for understanding the halo distribution
and formation.
  
CFHTLS is one of the new generation surveys, with
similar studies begining
at SUBARU, soon at ESO, with the VST, later with VISTA, also the NOAO deep survey
\footnote{http://www.noao.edu/noao/noaodeep/},
Dark Matter Telescope \footnote{http://www.dmtelescope.org/dark\_home.html},
and the PAN-STARRS \footnote{http://www.ifa.hawaii.edu/pan-starrs/}.
 Beyond 2005, space based dark energy/matter probes like SNAP 
  appear as a kind of final achievement.  In principle, SNAP can 
provide deep images, accurate photometric redshift, a large
  field of view and
outstanding image quality one expect for cosmic shear. A dark
matter space telescope, entirely dedicated to cosmic shear observations,
might also be an interesting option: it could be a 'small'
telescope (therefore fairly {\it cheap}) that could observe the
shear over all the sky. One could then 'see' the dark matter
everywhere!

Cosmic shear data are optimized when they are used
together with other surveys, like Boomerang, CBI, DASI, WMAP of 
  Planck CMB experiments, SNIa surveys, or galaxy surveys (2dF, SDSS).  
The first tentative  recently done by Contaldi,  Hoekstra \&
Lewis (2003) shows that tight constraints can really be 
  expected in the future. Likewise,  
  by using cosmic magnification instead of 
  cosmic shear on the  100, 000 SDSS quasars,  M\'enard \& Bartelmann
   (2002) have shown the cross-correlations between  the 
 foreground galaxy distribution and the quasar sample is also useful to 
  explore the properties of the biasing.  
    In principle magnification bias in the
SDSS quasar sample can provide similar constrains as cosmic
shear. Yet, this is a widely unexplored road.

This review shows that the Aussois winter school has been organized at a key 
period, between the first and second generations of cosmic shear surveys.
The first period, from 1999 to 2003,
demonstrated cosmic shear can be detected and exploited for a lot of 
cosmological questions. We now enter the next generation surveys,
which will end around 2010. These are large ground
based surveys (like the CFHTLS-weak lensing) which will fully exploit
the new windows opened by the first surveys.
Then we will enter the third (last?) period
with extensive space observations, probably after 2010, like
SNAP, which will permit to do precision cosmology, and maybe to close the
subject.

\acknowledgements
We are grateful to David Valls-Gabaud and Jean Paul Kneib, who organised a
very exciting lensing school that closes the 
  activivites of the LENSNET network. We would like to thank Matthias
  Bartelmann, Karim Benabed,
  Emmanuel Bertin,
Francis Bernardeau, David Bacon, Dick Bond, Carlo Contaldi,
Takashi Hamana,
Henk Hoekstra, Bhuvnesh Jain,
  Brice M\'enard, Ue-Li Pen, Dmitri Pogosyan, Simon Prunet, Alexandre 
  R\'efr\'egier, 
Peter Schneider and Ismael Tereno for regular stimulating discussions.  
This work was supported by the TMR Network
``Gravitational  Lensing: New Constraints on
Cosmology and the Distribution of Dark Matter'' (LENSNET) of the EC under contract
No. ERBFMRX-CT97-0172.

\label{page:last}
\end{document}